\documentclass[12pt]{iopart}

\usepackage{graphicx}
\usepackage{subfigure}
\usepackage{color}

\begin{document}

\title[]{Limitations on the quantum non-Gaussian characteristic of Schr\"{o}dinger kitten state generation}

\author{Hongbin Song$^{1}$, Katanya B. Kuntz$^{1,2}$, and Elanor H. Huntington$^{1}$}

\address{$^{1}$ School of Engineering and Information Technology, University of New South Wales Canberra, ACT 2610, Australia \\
$^{2}$ Center for Quantum Computation and Communication Technology, University of New South Wales Canberra, ACT 2610, Australia
}
\ead{hongbin.song@yahoo.com.au}
\begin{abstract}
A quantitative analysis is conducted on the impacts of experimental imperfections in the input state, the detector properties, and their interactions on photon-subtracted squeezed vacuum states in terms of a quantum non-Gaussian character witness and Wigner function. Limitations of the non-classicality and quantum non-Gaussian characteristic of Schr\"{o}dinger kitten states are identified and addressed. The detrimental effects of a photon-number detector on the generation of odd Schr\"{o}dinger kitten state at near-infrared wavelengths ($\sim$ 860 nm) and telecommunication wavelengths ($\sim$ 1550 nm) are presented and analysed. This analysis demonstrates that the high dark count probability of telecommunication-wavelength photon-number detectors significantly undermines the negativity of the Wigner function in Schr\"{o}dinger kitten state generation experiments. For a one-photon-subtracted squeezed vacuum state at $\sim$ 1550 nm, an APD-based photon-number-resolving detector provides no significant advantage over a non-photon-number-resolving detector when imperfections, such as dark count probability and inefficiency, are taken into account.

\end{abstract}


\maketitle

\section{Introduction:}
Non-Gaussian  operations on states have attracted intense interest in quantum continuous variable (QCV) information processing as they provide significant advantages for universal quantum computing \cite{PRL1999}, quantum teleportation \cite{sci2011}, entanglement distillation \cite{PRA2001,NP2010}, high-precision measurement \cite{PRA2002}, and proposed loophole-free tests of Bell's inequalities \cite{PRA2003}. Nowadays, two main categories of non-Gaussian and non-classical optical quantum states with negative-valued Wigner functions, such as Fock states \cite{PRL2006,METERO2006,OE2011} and photon-subtracted squeezed states \cite{SCI2006,OE2007,PRA2006604,PRA2010}, have been experimentally generated based on parametric down-conversion (PDC) in nonlinear crystals followed by photon number detection. In contrast to non-degenerate PDC for Fock state generation, degenerate PDC is used to generate photon-subtracted squeezed states. A small fraction of the squeezed vacuum beam is tapped off via a beam splitter and guided into a photon-number detector. The tapped-off light is used as a trigger to condition the remaining beam into a photon-subtracted squeezed vacuum state \cite{OE2007}. The projected state is referred to as an optical "Schr\"{o}dinger  kitten" as it closely approximates an optical Schr\"{o}dinger  cat state with a small amplitude \cite{SCI2006,OE2007}.

Photon-subtracted squeezed vacuum states with negative Wigner functions have been successfully demonstrated using Ti:Sapphire lasers at wavelengths around 860 nm and nonlinear crystals such as potassium niobate (KNbO3) and periodically-poled KTiOPO4 (PPKTP) \cite{SCI2006,OE2007,PRA2006604,PRA2010}. However, to our knowledge, a negative-valued Wigner function for Schr\"{o}dinger kitten state at telecommunication wavelengths $\sim$ 1550 nm is yet to be experimentally demonstrated \cite{NPHY2010}. We hypothesize that the principal difference in Schr\"{o}dinger kitten states generation between 860 nm and 1550 nm lies in the performance of photon-number detectors used in state preparation, which may undermine the negativity of the Wigner functions. Non-Gaussian states at telecommunication wavelengths are indispensable for secure optical quantum telecommunication due to their low loss in optical fibres. Therefore, it is imperative to develop a model to analyse the properties of Schr\"{o}dinger kitten states, and improve the experimental design based on the theory.

  Historically, negativity in the Wigner function has been the standard criterion to identify whether a state generated from an experiment is non-classical \cite{JOB2004,PhysRep1984}. However, the negative Wigner function of a quantum state generated from an experiment easily
degrades and becomes positive since the quantum state is fragile to any loss before it arrives at the verifying detector. Therefore, it is not suitable to solely rely on this criterion to characterise the non-classicality. To resolve this conflict, Jezek {\it et al} \cite{PRL2011602,ARX2012} proposed a quantum non-Gaussian character witness to verify states with positive Wigner functions that cannot be prepared by merely using Gaussian states and operations.

Dakna {\it et al} proposed the concept of "conditional measurement" based on a lossless beam splitter to generate a Schr\"{o}dinger  kitten state by subtracting photons from a squeezed vacuum state. They developed a model by taking the squeezing level, beam splitter transmission, photon-number detector inefficiency, and non-photon-number-resolving ability into account \cite{PRA1997,ARX1998}. The impacts of these factors on the Wigner function and non-classicality of a state were investigated by Olivares {\it et al} \cite{JOBQSO2005}. Kim {\it et al} analysed the necessary conditions to obtain a negative Wigner function for a realistic case including the input as a mixed state, threshold detection, inefficient homodyne detection, and mode purity in the subtraction path \cite{PRA2005}. The dark count influence of an on-off photon-number detector on the Wigner function of a Schr\"{o}dinger kitten state was considered by Suzuki {\it et al} \cite{OC2006}, and all of these imperfections were incorporated into a model developed by Gerrits {\it et al} \cite{PRA2010}. However, a quantitative analysis has not been conducted of the impacts all these experimental imperfections have on the negativity of the Wigner function and on the quantum non-Gaussian character witness of Schr\"{o}dinger kittens. Particularly, a thorough comparison between kitten state generation at $\sim$ 860 nm and generation at $\sim$ 1550 nm is yet to be explored.

In this paper, we quantitatively analyse the impacts of experimental parameters involving the impurity of the input state, inefficiency of the photon-number detector, dark count probability, non-photon-number-resolving ability, mode purity, and the homodyne detector inefficiency on Schr\"{o}dinger kittens generation in terms of the quantum non-Gaussian character witness proposed in references \cite{PRL2011602,ARX2012} and the Wigner function value at the origin, W(0,0). The comparisons between Schr\"{o}dinger kittens at near-infrared wavelengths ($\sim$ 860 nm) and telecommunication wavelengths ($\sim$ 1550 nm) are discussed, and principal limitations of state generation at telecommunication wavelengths are identified.

The paper is organized as follows. In Section 2, a Schr\"{o}dinger kitten state model is derived covering all possible experimental imperfections based on conditional measurement. In Section 3, the quantum non-Gaussian character witness introduced in references \cite{PRL2011602,ARX2012} is described. In Section 4, we present the physical mechanism for each experimental imperfection and quantitatively analyse the effects of these impacts on Schr\"{o}dinger kitten state features. Concluding remarks are given in Section 5.

\section{Schr\"{o}dinger  kitten state generation based on conditional measurement}

\begin{figure}[htbp]
  \centering
  \includegraphics[width=14cm]{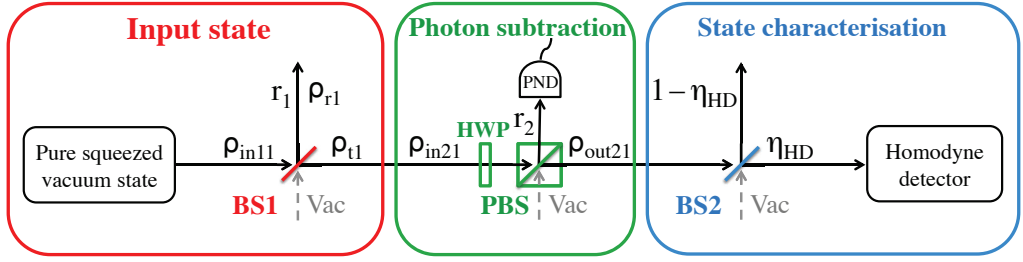}
  \caption{Theoretical model for Schr\"{o}dinger kitten state generation with experimental imperfections. $\rho_{in21}$ = $\rho_{t1}$, $r_{1}$: reflectivity of BS1, $t_{1}$: tansmission of BS1, $r_{2}$: reflectivity of PBS, $\rho_{in11}$: input state density matrix of BS1, $\rho_{t1}$: transmission state density matrix of BS1, $\rho_{r1}$: reflected state density matrix of BS1, $\rho_{in21}$: input state density matrix of PBS, $\rho_{out21}$: output state density matrix of PBS, $\eta_{HD}$: homodyne detection efficiency, BS1: beam splitter for modelling the impurity of the squeezed vacuum state, HWP: half-wave plate, PBS: polarization beam splitter, PND: photon-number detector, Vac: Vacuum state,  BS2: beam splitter for modelling the inefficiency of the homodyne detector.The HWP and the PBS comprise a 'magic' reflector. The HWP and the PBS comprise a 'magic' reflector.}
  \label{fig1schematicdigram}
\end{figure}

A schematic diagram of a theoretical model for Schr\"{o}dinger  kitten state generation with experimental imperfections is shown in figure \ref{fig1schematicdigram} \cite{PRA2010}. The model includes three parts: input state, photon subtraction, and state characterisation. The photon subtraction unit is composed of a `magic' reflector that is arbitrarily tuneable via a half-wave plate and a polarization beam splitter \cite{OE2007,PRA2010}. Ideally, when an even (odd) number of photons are subtracted from a pure squeezed vacuum state, an even (odd) kitten with a negative Wigner function can be obtained. However, numerous factors can undermine the ability of such an experiment to produce a Wigner function with negativity. These factors include optical elements related to the experiment, such as the impurity of the input squeezed vacuum state, mode impurity before the photon-number detector, and inefficiency of the homodyne detector used to characterise the quantum state. Imperfections in the photon-number detector can further degrade the prepared state. These imperfections involve a high dark count probability, low quantum efficiency, and the non-photon-number-resolving ability of some detectors. Therefore, a quantitative analysis of all these imperfections can shed light on the practical generation of Schr\"{o}dinger kitten states, particularly with regards to experiments at telecommunication wavelengths.

\subsection{Impure input state}
Parametric down-conversion based on a second-order nonlinear process in nonlinear materials is an effective approach to generate squeezed vacuum states. Ideally, a squeezed vacuum state consists of a photon number distribution with only even photon numbers. Impurity in the state contaminates the photon number distribution with odd photon number probabilities. This impurity can be equivalent to loss in a pure squeezed vacuum state, and can be described as a pure squeezed vacuum state followed by a beam splitter as shown in figure \ref{fig1schematicdigram}.

A pure squeezed vacuum state with a squeezing angle of zero degree can be expressed as \cite{LP2002}:
\begin{equation}
  \widehat{S}(\xi)|0\rangle=\sum_{n=0}^{\infty}\alpha_{2n}|2n\rangle,
                                                         \label{equ1}
\end{equation}
where

\begin{equation}
     \alpha_{2n}=\frac{1}{\sqrt{\\cosh\xi}}\frac{\sqrt{(2n)!}(-\\\\\tanh\xi)^{n}}{2^{n}n!}.\label{equ2}
\end{equation}

The input state to BS1 is written as
\begin{equation}
 |\Psi_{in1}\rangle=\widehat{S}(\xi)|0\rangle_{in11}|0\rangle_{1u}.\label{equ3}
                \end{equation}

We have an output state as
\begin{eqnarray}
|\Psi_{out11}\rangle &=\widehat{U}|\Psi_{in1}\rangle\nonumber
                   \nonumber\\&=\sum_{n=0}^{\infty}\alpha_{2n}\frac{\widehat{a}_{in11}^{+2n}}{\sqrt{(2n)!}}(\sqrt{r_{1}}\widehat{a}_{r1}^{+}+\sqrt{t_{1}}\widehat{a}_{t1}^{+})^{2n}|0\rangle_{r1}|0\rangle_{t1}
                      \nonumber\\&=\sum_{n=0}^{\infty}\sum_{k=0}^{2n}\alpha_{2n}\sqrt{\frac{(2n)!}{k!(2n-k)!}}r_{1}^{\frac{k}{2}}t_{1}^{\frac{2n-k}{2}}|k\rangle_{r1}|2n-k\rangle_{t1},\label{equ4}
\end{eqnarray}
where $\widehat{U}$ is the unitary operator of the beam splitter.

The transmitted density matrix after BS1 (i.e. the impure squeezed vacuum state) can be obtained by tracing the output density matrix over {\it $r_{1}$} \cite{book2008},

\begin{eqnarray}
\widehat{\rho}_{t_{1}} &=Tr_{r_{1}}[{\widehat{\rho}_{out1}}]\nonumber\\
                     &=\sum_{n,b=0}^{\infty}\sum_{k=0}^{2min(n,b)}\sqrt{\frac{(2n)!(2b)!}{(2n-k)!(2b-k)!}}\frac{\alpha_{2n}\alpha_{2b}r_{1}^{k}t_{1}^{n+b-k}}{k!}|2n-k\rangle\langle2b-k|,
                     \label{equ5}
\end{eqnarray}
where%
    \begin{equation}
  \widehat{\rho}_{out11}=|\Psi_{out11}\rangle\langle\Psi_{out11}|.\label{equ6}
\end{equation}

\subsection{Conditional measurement based on a lossless beam splitter}

According to the conditional beam splitter operator in reference \cite{ARX1998}, we have
     \begin{equation}
      |\Phi\rangle_{out21}=\frac{\widehat{Y}|\Phi_{in21}\rangle}{\|\widehat{Y}|\Phi_{in21}\rangle\|},\label{equ7}
\end{equation}

    where $ \|\|$ denotes the magnitude of a state vector and the non-unitary beam splitter operator is
   \begin{eqnarray}
        \widehat{Y} &=\langle\Phi_{out22}|\widehat{U}|\Phi_{in22}\rangle \nonumber\\&=\frac{t_{2}^{\frac{\widehat{n}_{1}}{2}}t_{2}^{\frac{m}{2}}{(-r_{2}^{*}})^{m}(\widehat{a}_{21})^{m}}{\sqrt{m!}},\label{equ8}
  \end{eqnarray}

thus
         \begin{eqnarray}
       \widehat{\rho}_{out21}(m)&=|\Phi_{out21}\rangle\langle\Phi_{out21}|
        \nonumber\\&=\frac{\frac{t_{2}^{m}}{m!}|r_{2}|^{m}t_{2}^{\frac{\widehat{n}_{1}}{2}}\widehat{a}_{21}^{m}\widehat{\rho}_{in21}\widehat{a}_{21}^{+m}t_{2}^{\frac{\widehat{n}_{1}}{2}}}{Tr[\frac{t_{2}^{m}}{m!}|r_{2}|^{m}t_{2}^{\frac{\widehat{n}_{1}}{2}}\widehat{a}_{21}^{m}\widehat{\rho}_{in21}\widehat{a}_{21}^{+m}t_{2}^{\frac{\widehat{n}_{1}}{2}}]},\label{equ9}
         \end{eqnarray}
  where $\widehat{\rho}_{in21}$ $=$ $\widehat{\rho}_{t1}$ as shown in figure {\ref{fig1schematicdigram}} and expressed in (\ref{equ5}), and Tr[ ] denotes the trace of a matrix.

\subsection{ Schr\"{o}dinger  kitten state prepared with an ideal photon-number-resolving detector}

In the case of an ideal photon-number-resolving detector (PNRD) (i.e. no dark counts and the quantum efficiency is 100\%), we can obtain the projected state density matrix by substituting (\ref{equ5}) into (\ref{equ9}),
 \begin{equation}
 \widehat{\rho}_{out21}(m)=\frac{\rho_{PNRD}(m)}{Tr[\rho_{PNRD}(m)]}\label{equ10}
   \end{equation}

 where
 \begin{eqnarray}
\rho_{PNRD}(m)=\sum_{n=0}^{\infty}\sum_{b=0}^{\infty}\sum_{k=0}^{2min(n,b)-m}\frac{\alpha_{2n}\alpha_{2b}r_{1}^{k}r_{2}^{m}(t_{1}t_{2})^{n+b-k}}{m!k!}\nonumber\\\sqrt{\frac{(2n)!(2b)!}{(2n-k-m)!(2b-k-m)!}}|2n-k-m\rangle\langle2b-k-m|.\label{equ11}
\end{eqnarray}

However, an ideal PNRD is unavailable in practical experiments. Avalanche photodiodes (APDs) are usually used as photon-number detectors, where Si-APDs and InGaAs-APDs are used to detect near-infrared wavelengths ($\sim $ 860 nm) and  telecommunication wavelengths ($\sim$ 1550 nm), respectively. Therefore, it is imperative to consider all possible imperfections of the photon-number detector, including the dark count probability, quantum efficiency, and the non-photon-number-resolving ability, and implement a quantitative analysis on the impact of all these experimental imperfections on the resultant quantum state.

\subsection{Schr\"{o}dinger  kitten state prepared with an imperfect photon-number detector}
\subsubsection{Dark counts probability and quantum efficiency}

On the one hand, the existence of dark counts causes `false' clicks even if a photon is not actually subtracted. On the other hand, some actual clicks are missed due to the inefficiency of the detector. Therefore, {\it m}-click events may originate from {\it m-1}, {\it m-2},...0 or {\it m+1}, {\it m+2},... actual photons being subtracted. Consequently, the conditional state is a statistical mixture, which can be expressed as \cite{PRA2010,PRA1997}:

 \begin{equation}
   \widehat{\rho}_{IMPNRD}(m)=\sum_{k=0}^{\infty}Q(k|m)\widehat{\rho}_{out1}(k),\label{equ12}
\end{equation}

where $Q(k|m)$ is defined as the conditioned probability, with which {\it m} photons would have been subtracted, given that {\it k} photons are actually   detected  by the imperfect detector. According to the Bayes rule, we can obtain the conditional probability,
\begin{equation}
   Q(k|m)=\frac{P(m|k)S(k)}{P(m)},\label{equ13}
\end{equation}

where
         \begin{eqnarray}
         S(k)=\sum_{n=k}^{\infty}\sum_{l=0}^{\infty}\sum_{b=0}^{\infty}\sum_{s=0}^{2min(l,b)}\alpha_{2n}\alpha_{2b}r_{1}^{s}t_{1}^{l+b-s}r_{2}^{k}t_{2}^{n-k} \label{equ14}
         \nonumber\\\sqrt{\frac{(2l)!(2b)!}{(2l-s)!(2b-s)!}} \nonumber\\\frac{n!}{k!s!(n-k)!}\langle n|2l-s\rangle\langle2b-s|n\rangle
           \end{eqnarray}
         is the probability of {\it k} photons being subtracted, which is calculated based on reference \cite{JMO2004}.

\begin{equation}
         P(m|k)=\sum_{d=0}^{m}e^{-P_{dc}}\frac{(P_{dc})^d}{d!}\frac{k!\eta_{APD}^{m-d}(1-\eta_{APD})^{k-m+d}}{(m-d)!(k-m+d)!},\label{equ15}
           \end{equation}

where $P_{dc}$ and $\eta_{APD}$ are the dark count probability and quantum efficiency of the detector, respectively \cite{JMO2009}.

\subsubsection {Non-photon-number-resolving ability}
Most photon-number detectors used in experiments so far are on-off or non-photon-number-resolving detectors (NPNRDs) without the capability to  distinguish the specific number of detected photons. Different from PNRD, NPNRD accepts {\it k} clicks even though the actual number of clicks can be larger than {\it k}. Thus we have \cite{PRA2010}

\begin{equation}
        \widehat{\rho}_{IMNPNRD}(m)=\sum_{k=m}^{\infty}\frac{Q(k)\widehat{\rho}_{outIMPNRD}(k)}{\sum_{k=m}^{\infty}Q(k)}.\label{equ16}
           \end{equation}

\subsection{Mode purity of subtracted photons}

Mode purity, {\it $s^{'}$}, is defined as the probability that the detected photons by the photon-number detector are mode matched to the local oscillator (LO) used in the kitten state characterisation via homodyne detection. As it is quite difficult to obtain a perfect mode purity, the detected density matrix of a projected state would be a mixed state consisting of the actual projected state and the unprojected state (i.e. the input state with loss). Therefore, we have \cite{PRA2005,PRL2004601}

          \begin{equation}
        \widehat{\rho}_{detect}=s^{'}*\widehat{\rho}_{projected}+(1-s^{'})*\widehat{\rho}_{input with loss}. \label{equ17}
           \end{equation}
\subsection{Schr\"{o}dinger  kitten state characterised by an inefficient homodyne detector}

Homodyne detection is a typical approach used to characterise the projected state. The homodyne detector efficiency is calculated as \cite{OE2007,PRA2006604,PRA2010}

\begin{equation}
        \eta_{HD}=\eta_{QE}*\eta_{t}*\zeta^{2},\label{equ18}
           \end{equation}
where
      $\eta_{QE}$ is the quantum efficiency of the two photodiodes in the homodyne detector, $\eta_{t}$ is the transmission coefficient from the `magic' reflector to the homodyne detector, and $\zeta$ is the visibility of interference fringes between the signal and LO, denoting the degree of mode matching. Therefore, the total efficiency, $\eta_{HD}$, quantifies various categories of loss. As show in figure \ref{fig1schematicdigram}, the homodyne detection inefficiency can be simulated by a lossless beam splitter before a perfect homodyne detector, and the density matrix measured with an inefficient homodyne detector is given by \cite{bookMQ},

      \begin{equation}
        \langle l|\widehat{\rho}_{detect}(\eta_{HD})|n\rangle=\sum_{k=0}^{\infty}B_{l+k,l}(\eta)B_{n+k,n}(\eta)\langle l+k|\widehat{\rho}_{detect}|n+k\rangle \label{equ19}
           \end{equation}
in terms of the initial field density matrix $\widehat{\rho}_{detect}$, where

      \begin{equation}
       B_{l+k,l}(\eta)=\sqrt{\frac{(l+k)!}{k!l!}\eta_{HD}^{l}(1-\eta_{HD})^{k}}.\label{equ20}
           \end{equation}

\section{Character witness to identify the quantum non-Gaussian state and non-classical state}
Up until very recently, negativity in the Wigner function has been widely used as the criterion to identify the non-classicality of a state \cite{JOB2004}. However, for some non-classical quantum states, such as squeezed vacuum states, this criterion does not work because they possess positive Wigner functions. In addition, some heralded quantum states have positive Wigner functions that could not be prepared from Gaussian states and linear optical devices. Therefore, Jezek {\it et al} \cite{PRL2011602,ARX2012} proposed a non-classical and a quantum non-Gaussian witness. States beyond a convex set of stochastic mixture of coherent states are defined as non-classical states. Similarly, quantum non-Gaussian states are referred to as states beyond a convex set of stochastic mixture of Gaussian states \cite{PRL2011602,ARX2012}.

The quantum non-Gaussian character witness is based on Fock state basis, and is introduced as a linear combination of zero photon probability, $p_{0}$, and one photon probability, $p_{1}$, in the Fock state basis density matrix  \cite{PRL2011602,ARX2012},

\begin{equation}
        W(a)=ap_{0}+p_{1}
        \label{equ21newwitness}
           \end{equation}
where
     \begin{eqnarray}
        p_{0}=\frac{e^{-e^{r}sinh{r}}}{cosh{r}},
        \label{equ22}
           \end{eqnarray}

     \begin{eqnarray}
        p_{1}=\frac{e^{4r}-1}{4}\frac{e^{-e^{r}sinh{r}}}{cosh{r}^{3}},
        \label{equ23}
           \end{eqnarray}
 $a\in[0, 1]$  is a dimensionless number and $r\in[0,\infty)$ is the squeezing parameter. A quantum Gaussian boundary, $W_G(a)$, is defined as the maximum value of {\it W(a)} over {\it a} and {\it r}. The quantum Gaussian character witness value is defined as {\it W(a)} - $W_G(a)$. If this witness value is larger than 0, then the state is a quantum non-Gaussian state. For quantum states related to squeezed states, such as squeezed single photon states or Schr\"{o}dinger  kitten states from photon-subtracted squeezed states, the quantum non-Gaussian character witness is generalized by an anti-squeezing operation \cite{ARX2012},

                 \begin{equation}
        W(a,s)=ap_{0}(s)+p_{1}(s)
        \label{equ24}
           \end{equation}

           where
        \begin{equation}
        p_{n}(s)=\langle n|S^{+}(s)\widehat{\rho} S(s)|n\rangle
        \label{equ25}
           \end{equation}

         $S^{+}$ and {\it S} are the anti-squeezing and squeezing operators, respectively, and $\widehat{\rho}$ corresponds to the density matrix of the state in the Fock state basis. The quantum non-Gaussian character witness value is defined as $W(a,s)-W_G(a)$ in the following sections.

Equivalently, a classical boundary is defined as the maximum value of

      \begin{equation}
        W_{cl}(a)=ap_{0}+p_{1}
        \label{equ21newwitness}
           \end{equation}

over {\it a}, where
   \begin{eqnarray}
        p_{0}=e^{-\overline{n}}\\
          p_{1}=\overline{n}e^{-\overline{n}},
        \label{equ22}
           \end{eqnarray}
      and $\overline{n} $ $\in$ $[0,\infty]$ is the mean photon number.

Therefore, it is easy to identify the quantum non-Gaussian or non-classical characteristic of a state via its density matrix.

\section{Quantitative analysis of experimental imperfections impact on Schr\"{o}dinger kitten state generation}

\subsection{Estimation of pure input state level and input state impurity}
Generally, squeezing value is referred to as the noise variance of a squeezed state, which is related to the degree of squeezing $\xi$ in (\ref{equ1}) by

\begin{equation}
        V_{sqz}=cosh(2\xi)-sinh(2\xi),
        \label{equ26}
           \end{equation}

for a squeezed state with a squeezing angle of {\it zero} degree. Here, we define the squeezing level as the base 10 logarithm of squeezing (i.e. the noise variance in dB).

The variance of a pure squeezed state, $V_{0}$, and the impurity, $r_{1}$, are experimentally determined by the measured squeezing, $V_{sqz}$, anti-squeezing, $V_{a-sqz}$, and the corresponding homodyne detector efficiency, $\eta_{HD}$ \cite{PRA2010},

\begin{equation}
        V_{0}=\frac{1-V_{sqz}}{V_{a-sqz}-1}
        \label{equ26}
           \end{equation}

\begin{equation}
        r_{1}=\frac{\eta_{HD}(2-V_{sqz}-V_{a-sqz})-(1-V_{sqz})(1-V_{a-sqz})}{(2-V_{sqz}-V_{a-sqz})\eta_{HD}}.
        \label{equ27}
           \end{equation}

A typical squeezed vacuum state after the 'magic' reflector in a Schr\"{o}dinger kitten state generation experiment as shown in figure \ref{fig1schematicdigram} can be obtained with $r_{2}$ = 0.08, squeezing of $V_{sqz}$ = 0.661 ($-$1.8 dB), anti-squeezing of $V_{a-sqz}$ = 1.995 (+3 dB), and a homodyne detection efficiency of  $\eta_{HD}$ = 68\%. According to (\ref{equ26}) and (\ref{equ27}), the corresponding pure squeezing and impurity are $V_{0}$ = 0.341 ($-$4.67 dB) and $r_{total}$ = 0.2438 ($r_{total}$ are the total impurity caused by $r_{1}$ and $r_{2}$), respectively. By taking $r_{2}$ into account, we actually have $r_{1}$ = 0.1771.

\subsection{Quantum non-Gaussian character witness for an impure squeezed state and Schr\"{o}dinger kitten state}

 Based on the model developed in Section 2, a one-photon-subtracted impure squeezed vacuum state ($V_{0}$ = -4.67 dB and impurity of $r_{1}$ = 0.1771 ) prepared with $r_{2}$ = 0.08, a non-photon-number-resolving detector with $P_{dc}$ = $1\times10^{-4}$, and $\eta_{APD}$ = 5\% was constructed.

\begin{figure} [htbp]
\centering
\subfigure[Photon number distribution]{
\label{fig:subfig:a} 
\includegraphics[height=2.4in,width=2.95in]{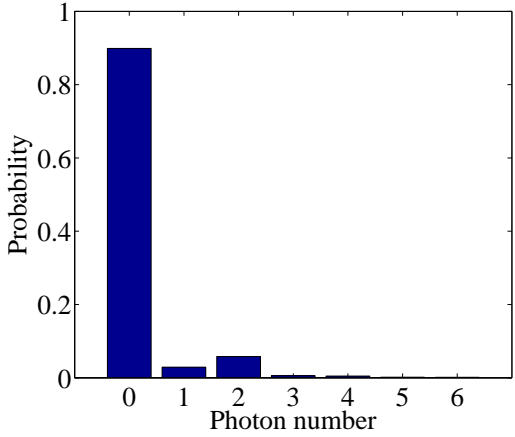}}
\hspace{0in}
\subfigure[Wigner function, W(0,0) = 0.2949]{ \label{fig:subfig:b} 
\includegraphics[height=2.4in,width=2.95in]{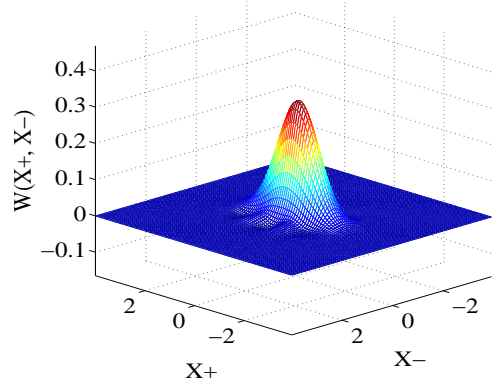}}
\caption{(a) Photon number distribution and (b) Wigner function of an impure squeezed vacuum state with $V_{0}$ = $-$4.67 dB and $r_{1}$ = 0.1771}
\label{fig2impuresqzvac} 
\end{figure}

\begin{figure} [hb]
\centering
\subfigure[Photon number distribution]{
\label{fig:subfig:a} 
\includegraphics[height=2.4in,width=2.95in]{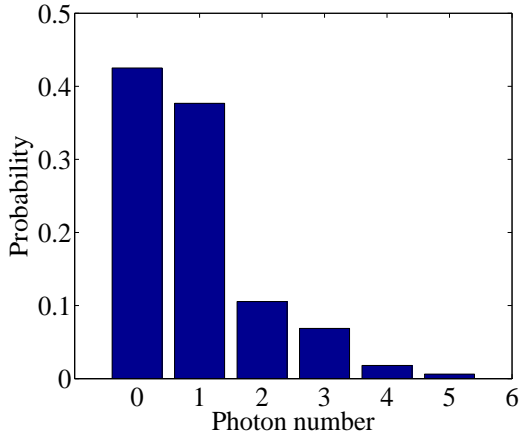}}
\hspace{0in}
\subfigure[Wigner function, W(0,0) = 0.0309]{ \label{fig:subfig:b} 
\includegraphics[height=2.4in,width=2.95in]{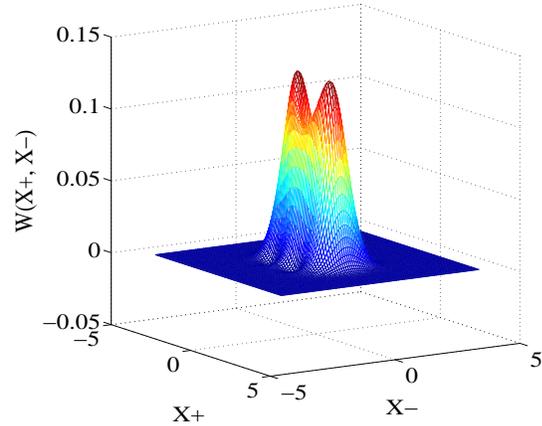}}
\caption{(a) Photon number distribution and (b) Wigner function of a projected state prepared with a non-photon-number-resolving detector and $V_{0}$ = $-$4.67 dB, $r_{1}$ = 0.1771, $r_{2}$ = 0.08, $P_{dc}$ = $1\times10^{-4}$, $\eta_{APD}$ = $5\%$, $\eta_{HD}$ = $100\%$, and mode purity = 1.}
\label{fig3projected} 
\end{figure}

\begin{figure}[htbp]
 \centering
 \includegraphics[height=2.4in,width=2.95in]{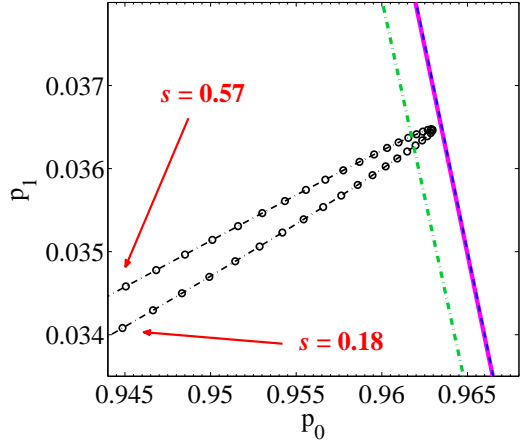}
\caption{({\it $p_{0}$}, {\it $p_{1}$}) trajectory of an impure squeezed vacuum state with anti-squeezing operation when anti-squeezing parameter {\it s} varies from 0.18 to 0.57 (black dot-dash line with circle). The dot-dash green and solid pink lines represent the classical boundary and the quantum Gaussian boundary, respectively. The dash blue line overlaps with the solid pink line and corresponds to the physical limit, $p_{0}+p_{1}=1$.}
 \label{fig4witnesssqzvac}
\end{figure}

The photon number distribution and Wigner function of the input impure squeezed vacuum state and the projected state are shown in figures \ref{fig2impuresqzvac} and \ref{fig3projected}. We can see that both states possess positive Wigner functions at the origin. However, the projected state is clearly a non-Gaussian state. Figure \ref{fig4witnesssqzvac} shows the impure squeezed vacuum state with anti-squeezing operation. We observed that the impure squeezed vacuum state is a non-classical but Gaussian state, as expected.

\begin{figure} [htbp]
\centering
\subfigure[]{
\label{fig:subfig:a} 
\includegraphics[height=2.4in,width=2.95in]{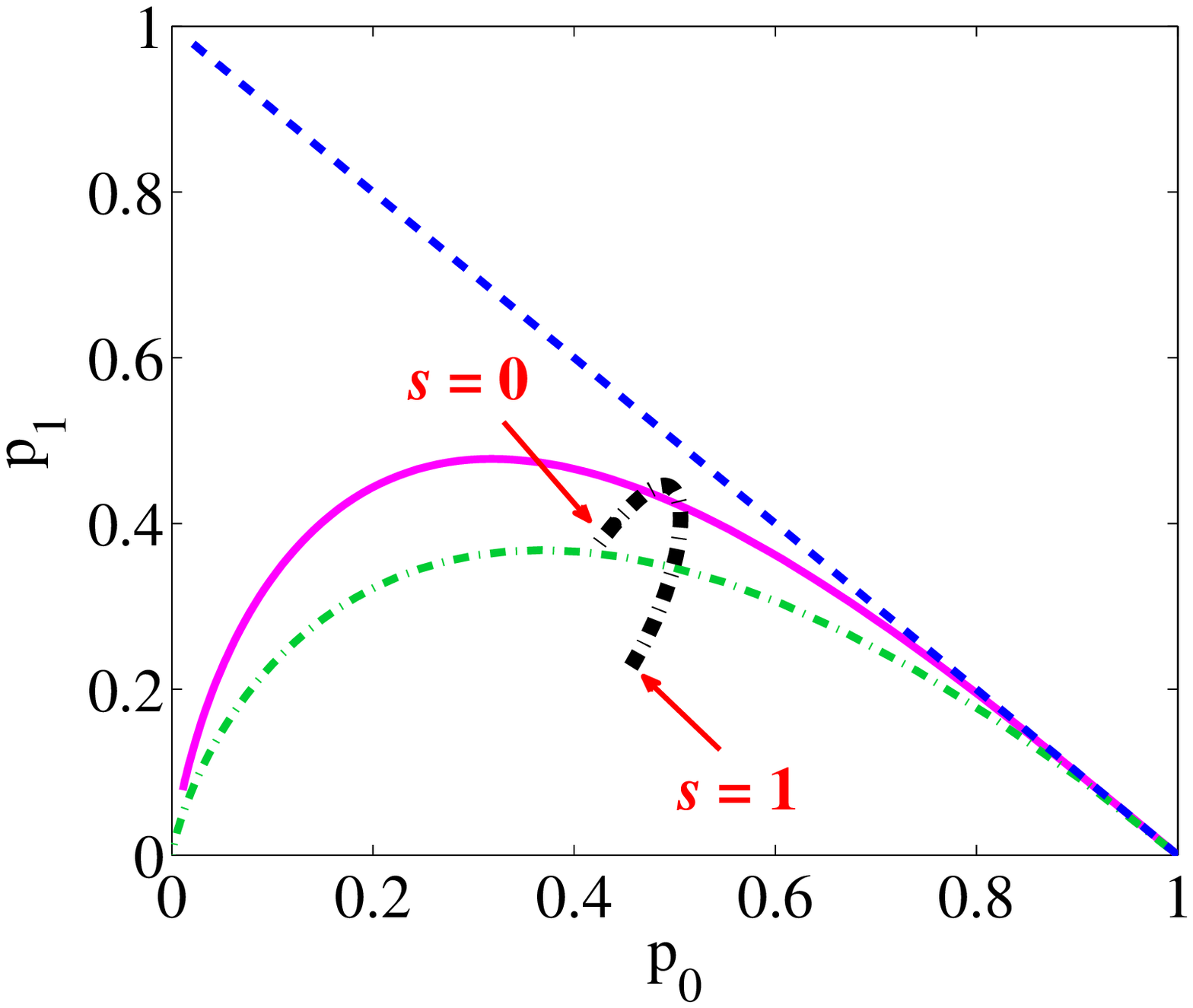}}
\hspace{0in}
\subfigure[]{ \label{fig:subfig:b} 
\includegraphics[height=2.4in,width=2.95in]{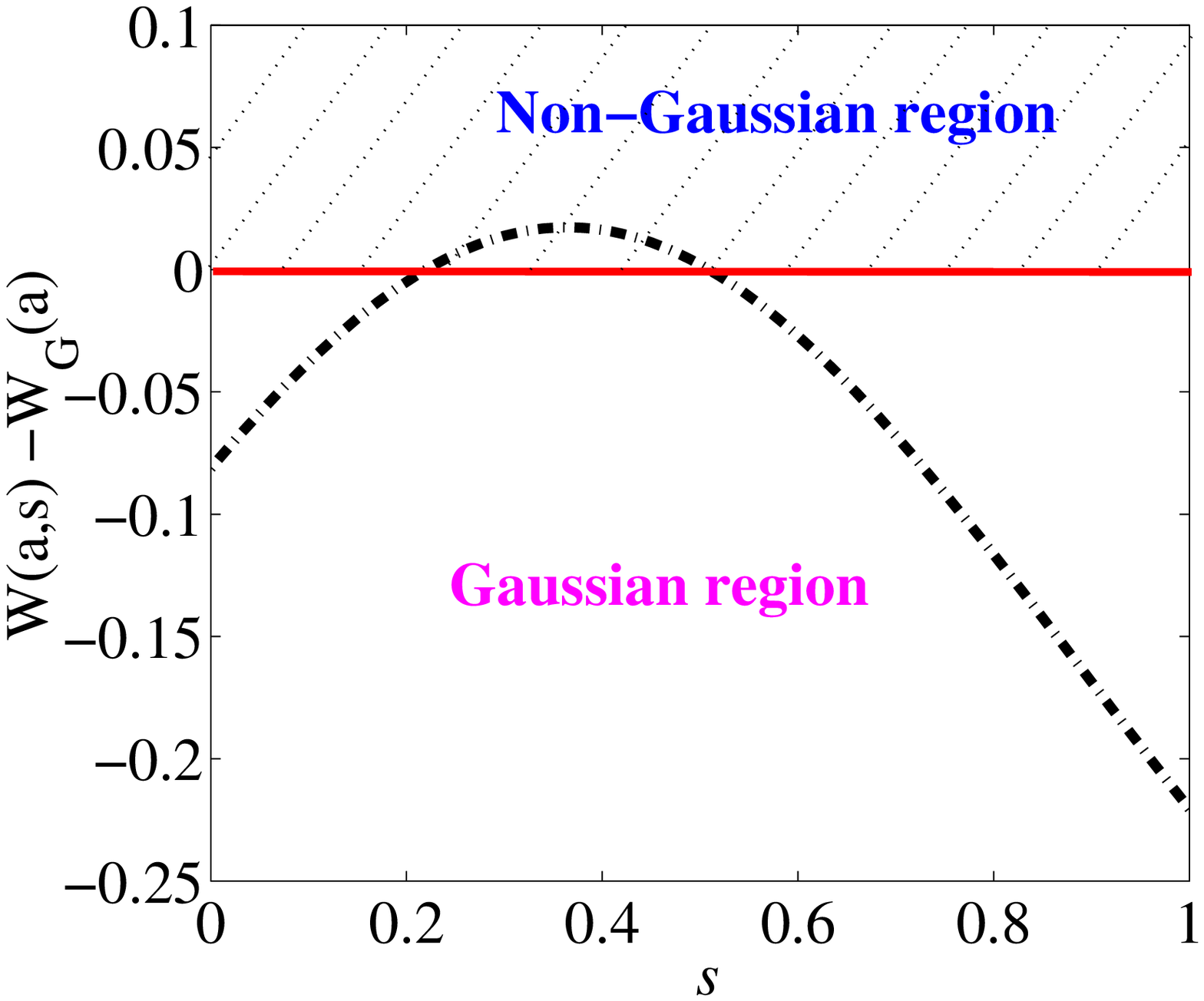}}
\caption{Characteristic identification of the projected state. (a) ({\it $p_{0}$}, {\it $p_{1}$}) trajectory of the projected state with anti-squeezing operation when anti-squeezing parameter {\it s} varies from 0 to 1 (dot-dash black line). The dot-dash green and solid pink lines represent the classical and quantum Gaussian boundaries. The dash blue line corresponds to the physical limit, $p_{0}+p_{1}=1$. (b) The optimal witness $W(a_{opt},s)-W_{G}(a_{opt})$ }
\label{fig5witnessprojected} 
\end{figure}

As indicated in figure \ref{fig5witnessprojected}(a), the one-photon projected Schr\"{o}dinger kitten state (corresponding to {\it s} = 0) is located beyond the classical state boundary but is within the quantum Gaussian state boundary. Apparently, this indicates that the state is non-classical and Gaussian. However, after the anti-squeezing operation, the kitten state crosses the quantum Gaussian state boundary, which implies the quantum non-Gaussian characteristic of the kitten state (i.e. it cannot be prepared by merely mixing Gaussian states). Moreover, the quantum non-Gaussian character witness implies that quantum non-Gaussian states possess strong non-classicality. Therefore, we confirm that the quantum non-Gaussian character witness proposed in \cite{ARX2012} demonstrates a powerful ability to identify non-classical and non-Gaussian quantum states.

\subsection{Physical mechanism underpinning each experimental imperfection}

The physical mechanism underpinning experimental imperfections are summarised in table \ref{tablephysicalmechanism}, from which we can categorise the underlying impacts as: 1) stochastic mixture of an {\it m}-photon-subtracted state with an {\it M}-photon-subtracted squeezed vacuum state, where ({\it M} $>$ {\it m}), 2) photon number redistribution of an ideal Schr\"{o}dinger kitten state, and 3) statistical mixture of the projected state with unprojected states.

\begin{table}[htbp]
\caption{\centering Physical mechanism underpinning experimental imperfections}
\begin{tabular*}
{\textwidth}{@{}l*{10}{@{\extracolsep{0pt plus0pt}}l}}
\br

\bf Impacts&\;\;\;\bf Experimental imperfections &\;\;\;\;\;\; \;\;\;\;\;\;\bf Physical mechanism\\ \hline
 1)&\;APD inefficiency, $\eta_{APD}$&\;{\it Statistically mixed m-photon-subtracted }\\ \cline{1-9}
 &\;Non-photon-number-resolving ability& \;{\it squeezed vacuum state with {\it M}-photon-} \\ \cline{1-9}
 &\;Squeezing level, $V_{0}$ &\;{\it ({\it M} $>$ {\it m}) subtracted state (i.e. {\it m}-click }\\ \cline{1-9}
 &\;Reflectivity, $r_{2}$,& \;{\it events actually result from {\it m+1}, {\it m+2} }\\ \cline{1-9}
  &\;& \;... {\it -photon subtraction)}\\ \cline{1-9} \hline
2)&\;Input state impurity, $r_{1}$  & \;{\it ${\it N}$ ({\it N} $>$ 0) photon in an ideal kitten } \\ \cline{1-9}
  &\;Homodyne detection inefficiency, $\eta_{HD}$ & \;{\it  state is redistributed to {\it N-1}, {\it N-2},}\\ \cline{1-9}
 &\; & \;{\it  {\it N-3},...0 }\\ \cline{1-9}
\hline
 3)&\;APD dark count, $P_{dc}$ &{\it Statistically mixed projected state with }\\ \cline{1-9}
 &\;Mode impurity, {\it s'} & \;{\it unprojected state (i.e. {\it m} click events } \\ \cline{1-9}
&\;&\;{\it may actual originate from {\it m-1}, {\it m-2}}\\   \cline{1-9}
&\;&\;{\it .... -photon subtraction)}\\   \cline{1-9}
\br
\label{tablephysicalmechanism}
\end{tabular*}
\end{table}

\begin{figure} [htpb]
\centering
\subfigure[IMPNRD $\eta_{APD} = 0.5$]{
\label{fig:subfig:a} 
\includegraphics[height=2.4in,width=2.95in]{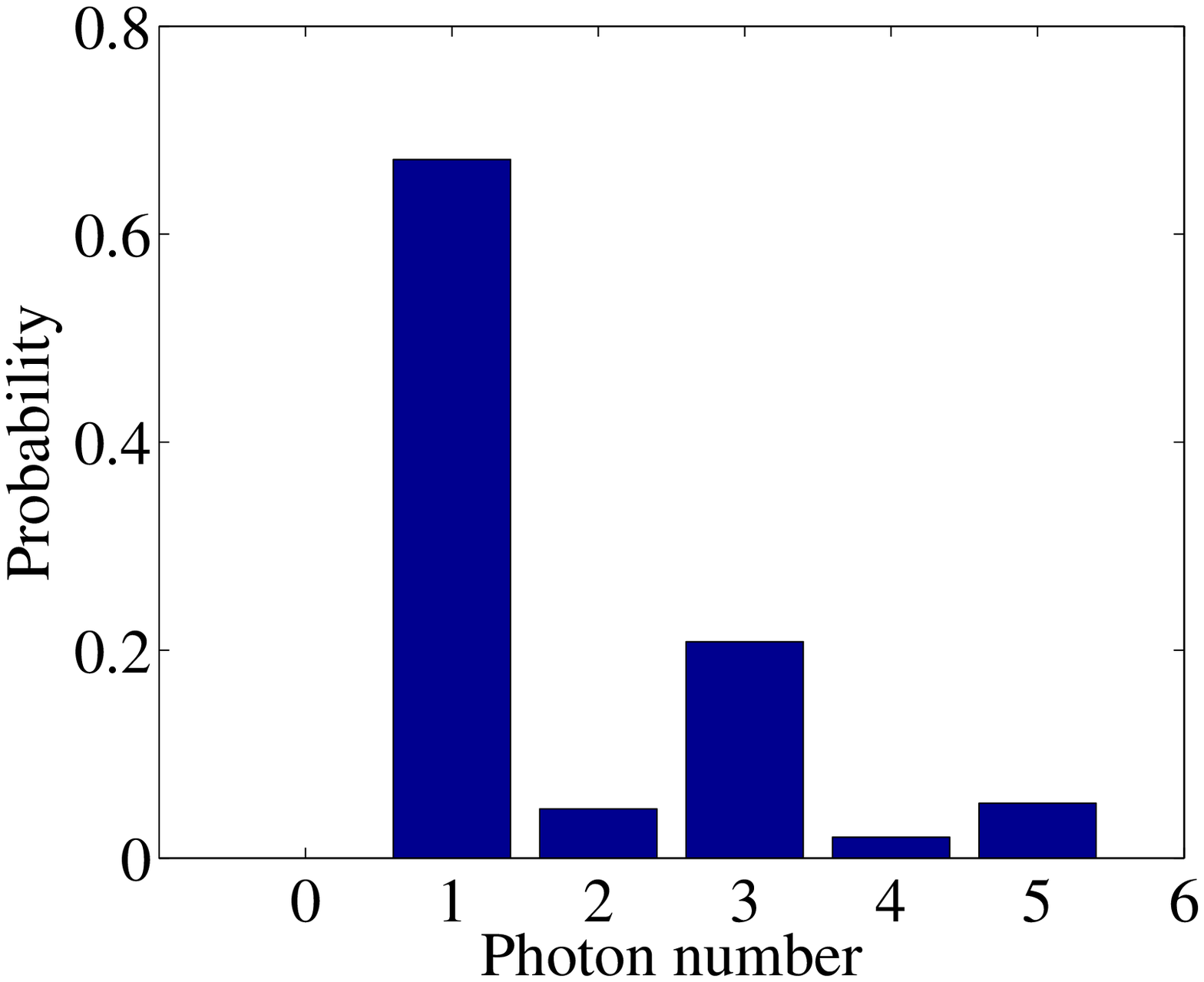}}
\hspace{0in}
\subfigure[NPNRD $\eta_{APD} = 1$]{\label{fig:subfig:b} 
\includegraphics[height=2.4in,width=2.95in]{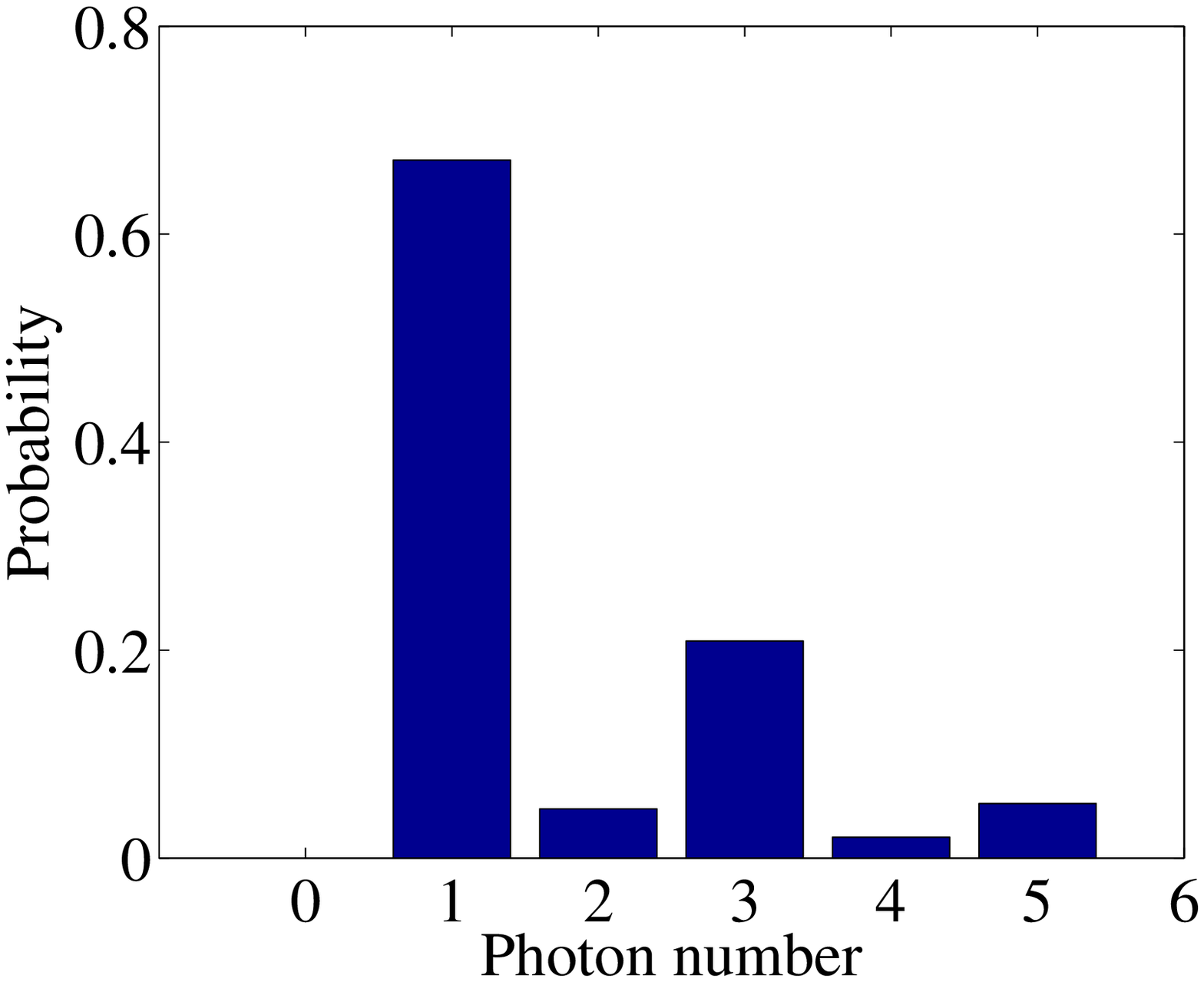}}
\caption{Photon number distribution of Schr\"{o}dinger kitten states prepared with (a) an inefficient APD and (b) a non-photon-number-resolving APD . In both cases, other parameters are: $V_{0}$ = $-$4.67 dB, $r_{1} $ = 0, $r_{2}$ = 0.08, $P_{dc}$ = 0, and mode purity = 1.}
\label{figetaapdnpnrd} 
\end{figure}

To illustrate the above statements, figure \ref{figetaapdnpnrd} shows the photon number distributions for states prepared with (a) an IMPNRD with ${\it \eta_{APD}}=50\%$, and ${\it P_{dc}=0}$, and (b) a perfect NPNRD with ${\it \eta_{APD}}=100\%$, and ${\it P_{dc}=0}$, where all other related parameters are set to be the same. The photon number distributions are quite similar, which implies that the advantage of photon number resolution dramatically reduces when the detection efficiency is low.

\begin{figure*} [htbp]
\centering
\subfigure[$r_{1}$ = 0.1771, $\eta_{HD}$ = 1]
{
\label{fig:subfig:a} 
\includegraphics[height=2.4in,width=2.95in]{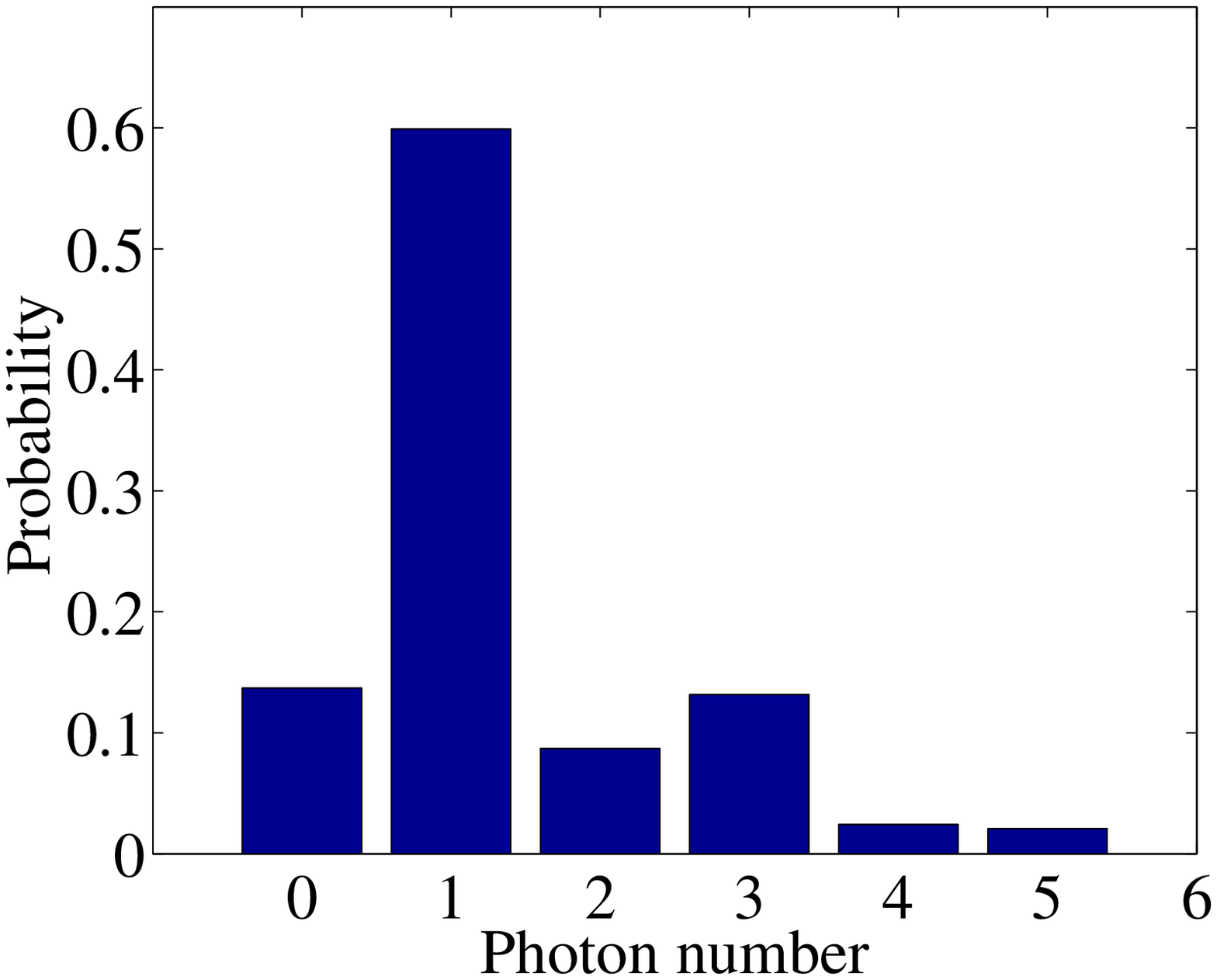}}
\hspace{0in}
\subfigure[$r_{1}$ = 0, $\eta_{HD}$ = 0.8]{ \label{fig:subfig:b} 
\includegraphics[height=2.4in,width=2.95in]{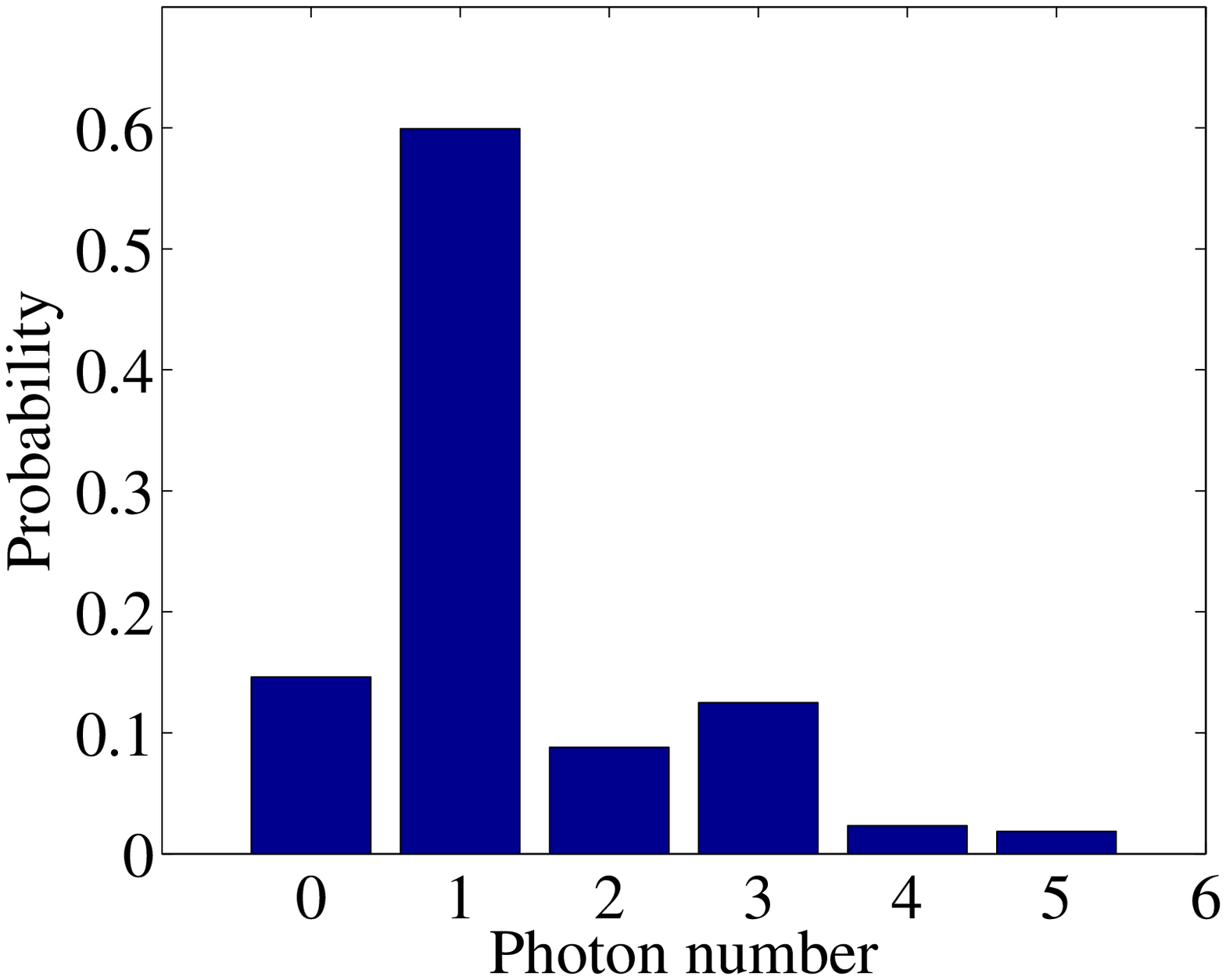}}
\caption{Photon number distributions of Schr\"{o}dinger kitten states with (a) an impure input state and (b) inefficient homodyne detection. In both cases, other parameters are: $V_{0}$ = $-$4.67 dB, $r_{2} = 0.08$, $P_{dc}$ = 0, $\eta_{APD}$ = 1, and mode purity = 1.}
\label{figr1etahd} 
\end{figure*}

\begin{figure} [htpb]
\centering
\subfigure[$P_{dc}$ = 0.005, mode purity = 1]{
\label{fig:subfig:a} 
\includegraphics[height=2.4in,width=2.95in]{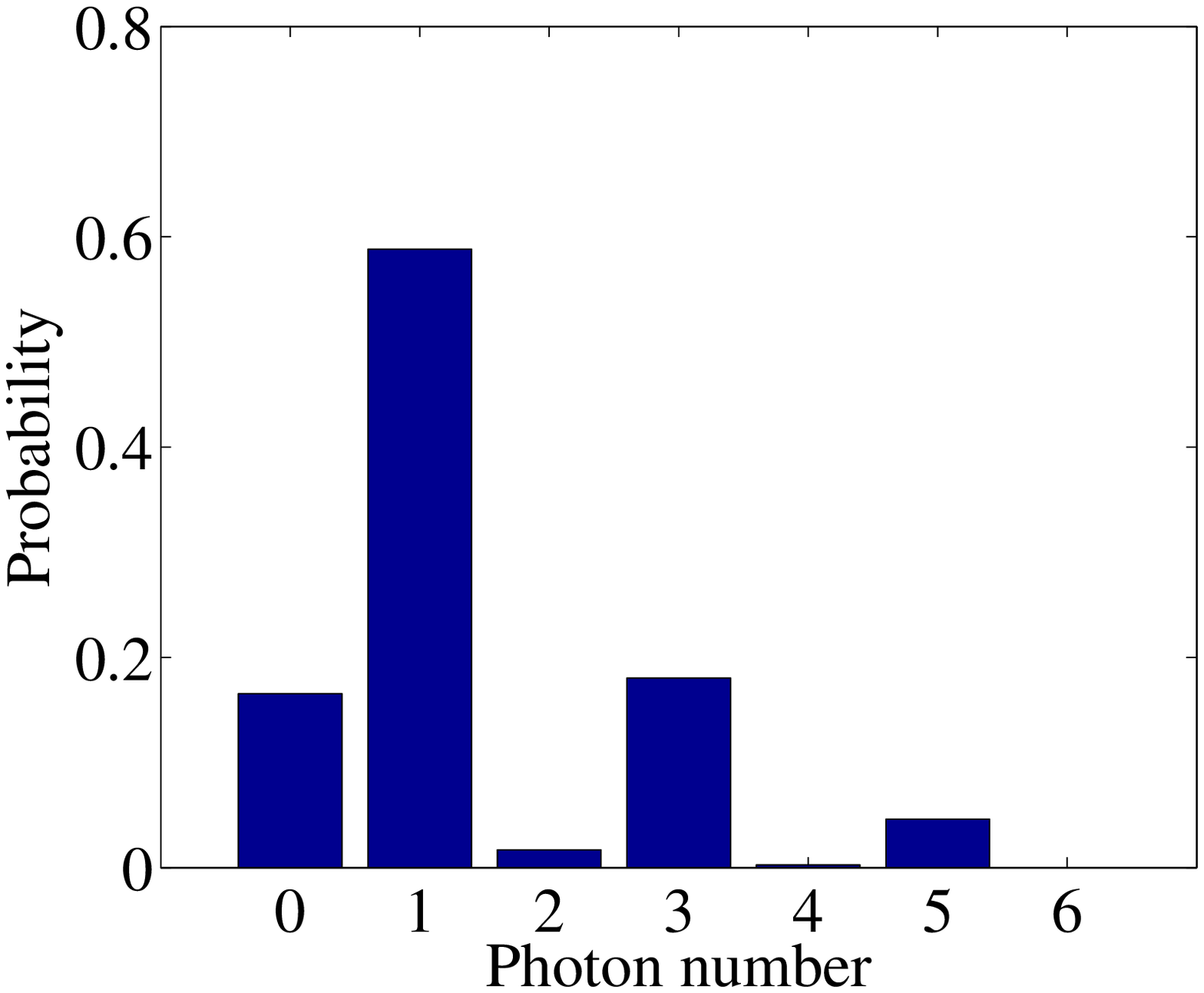}}
\hspace{0in}
\subfigure[$P_{dc}$ = 0, mode purity = 0.85]{ \label{fig:subfig:b} 
\includegraphics[height=2.4in,width=2.95in]{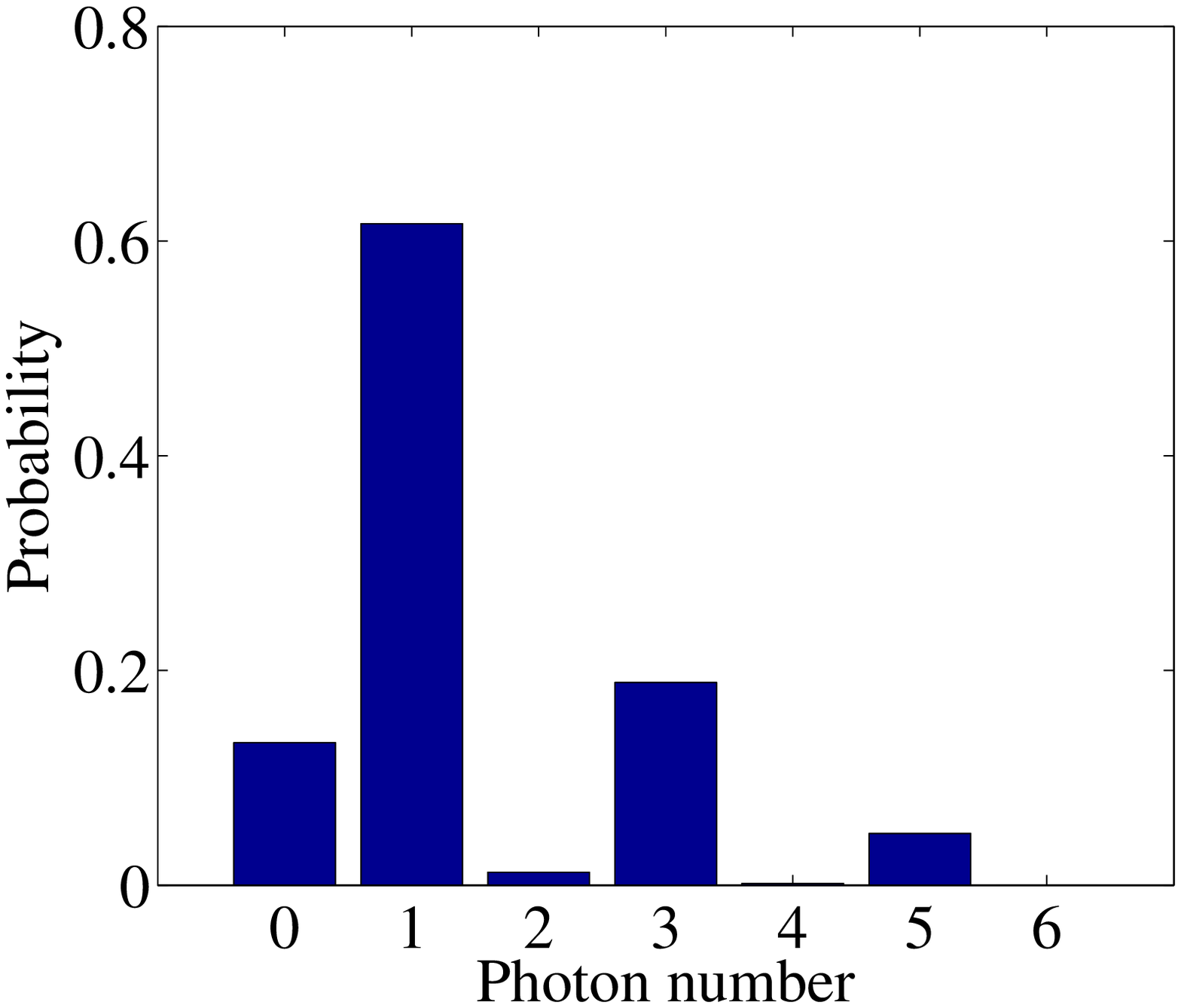}}
\caption{Photon number distributions of Schr\"{o}dinger kitten states prepared with (a) an APD with dark counts and (b) mode impurity. In both cases, other parameters are: $V_{0}$ = $-$4.67 dB, $r_{1} = 0.1771$, $r_{2}$ = 0.08, $\eta_{APD}$ = 1, and $\eta_{HD}$ = 1.}
\label{figpdcmodeimpurity} 
\end{figure}

As another example, figure \ref{figr1etahd} gives the photon number distributions for Schr\"{o}dinger kitten states prepared with (a) an input squeezed vacuum state with an impurity of 0.1771 and 100\% homodyne detection efficiency, and (b) an pure input squeezed vacuum state but 80\% homodyne detection efficiency, where other related parameters are the same. The similarity in photon number distributions indicates the equivalent quantitative effect of the input state impurity and homodyne detection inefficiency on the projected states. Furthermore, figure \ref{figpdcmodeimpurity}(a) and (b) verifies that a high dark count probability of an APD and mode impurity demonstrate equivalent impacts on kitten states.

\subsection{Dependency of the quantum non-Gaussian character witness and Wigner function at origin, W(0,0), on experimental imperfections}

Although superconducting transition edge sensors (TESs) with photon-number-resolving ability are available \cite{PRA2010,NPHY2010}, commercially available APDs are still widely used as photon-number detectors in Schr\"{o}dinger  kitten state generation experiments since cryogenic environments are required for TESs. Typical dark count probabilities and detection efficiencies of commercially available Si-APDs (for 860 nm from Perkin Elmer Ltd.) and InGaAs-APDs (for telecommunication wavelength from ID Quantique Ltd.) are listed in table \ref{tableAPD}.  Si-APDs perform better than InGaAs-APDS due to their lower dark count probabilities and higher detection efficiencies. A group of typical values for other related parameters are shown in table \ref{tableparamerters}. We will focus on discussing the impacts of experimental imperfections on the quantum non-Gaussian character witness value and Wigner function of a one-photon-subtracted vacuum state prepared with a Si-APD(SPCM-AQR-12) and an InGaAs-APD (id200).

\begin{table}[tp]
\caption{Comparison of Si-APDs \& InGaAs-APDs performance}
\footnotesize\rm
\begin{tabular*}{\textwidth}{@{}l*{15}{@{\extracolsep{0pt plus12pt}}l}}
\br
\bf Detector type&\bf Dark count probability($P_{dc}$) & \bf Quantum efficiency($\eta_{APD}$)\\
\mr
\verb"Si-APD(SPCM-AQR-12) "&$5\times10^{-6}$&$45\%$\\
\verb"Si-APD(SPCM-AQR-13) "&$2.5\times10^{-6}$&$45\%$\\
\verb"Si-APD(SPCM-AQR-14) "&$1\times10^{-6}$&$45\%$\\
\verb"Si-APD(SPCM-AQR-15) "&$5\times10^{-7}$&$45\%$\\
\verb"Si-APD(SPCM-AQR-16) "&$2.5\times10^{-7}$&$45\%$\\\hline

\verb"InGaAs-APD(id200)"&$1\times10^{-4}$&$10\%$\\
\verb"InGaAs-APD(id220)"&$1\times10^{-5}$&$10\%$\\
\verb"(under different settings)"&$2.5\times10^{-5}$&$15\%$\\
\verb                  &$5\times10^{-5}$&$20\%$\\
\br
\label{tableAPD}
\end{tabular*}
\end{table}

\begin{table}[b]
\caption{Typical experiment parameters used in the simulation}
\footnotesize\rm
\begin{tabular*}{\textwidth}{@{}l*{15}{@{\extracolsep{0pt plus12pt}}l}}
\br
\bf Parameter & \bf Typical value\\
\mr
\verb"Squeezing level",  $V_{0}$& $-$4.67 dB\\
\verb"Input state impurity", $r_{1}$& 0.1771 \\
\verb"Reflectivity", $r_{2}$ & 0.08\\
\verb"Mode purity", $s'$& 0.8\\
\verb"Homodyne detection efficiency", $\eta_{HD}$& 85\% \\
\br
\label{tableparamerters}
\end{tabular*}
\end{table}

\subsubsection{ Effects of squeezing level, $r_{2}$, and $\eta_{APD}$}

\begin{figure} [htbp]
\centering
\subfigure[InGaAs-APD]{
\label{fig:subfig:a} 
\includegraphics[height=2.4in,width=2.95in]{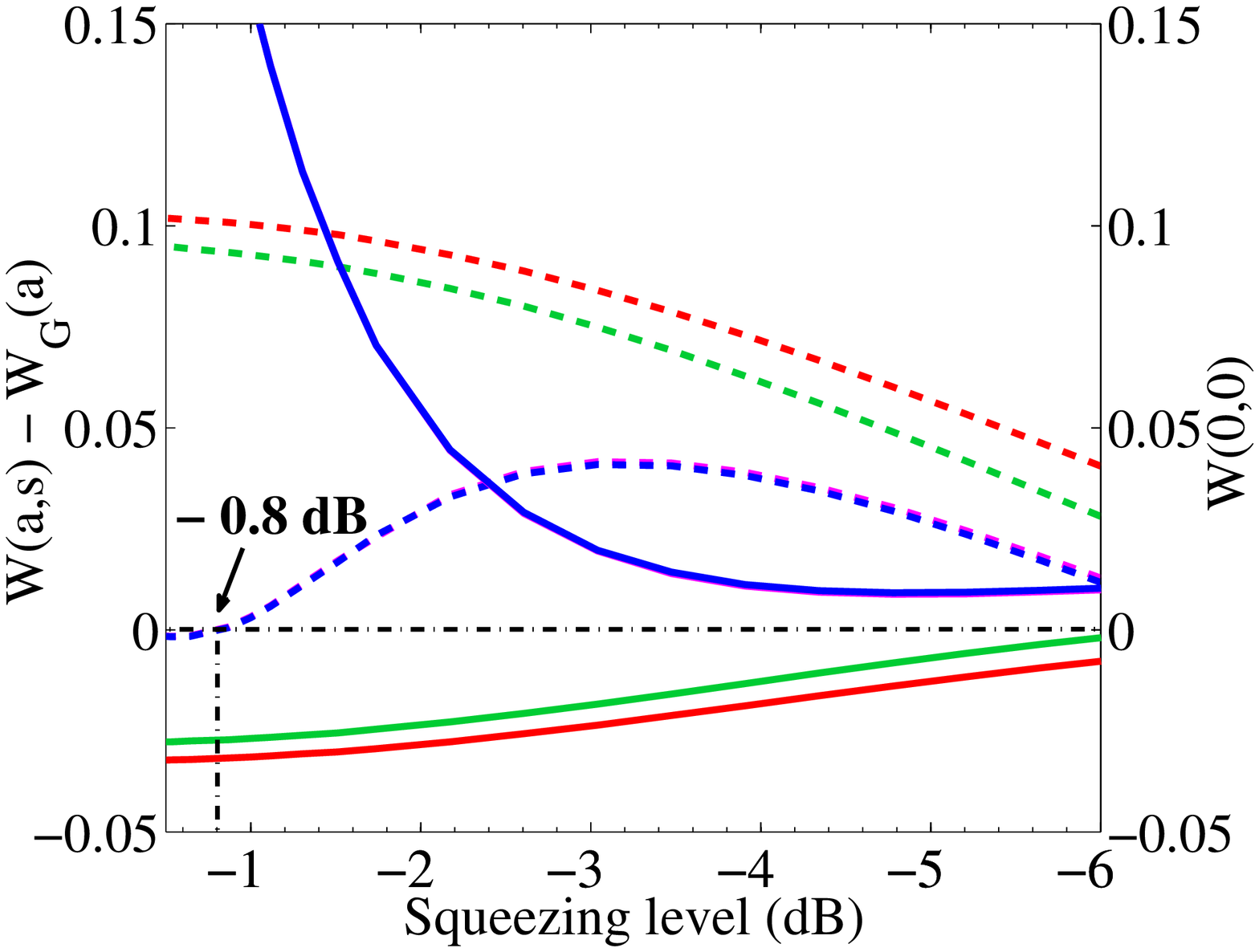}}
\hspace{0in}
\subfigure[Si-APD]{ \label{fig:subfig:b} 
\includegraphics[height=2.4in,width=2.95in]{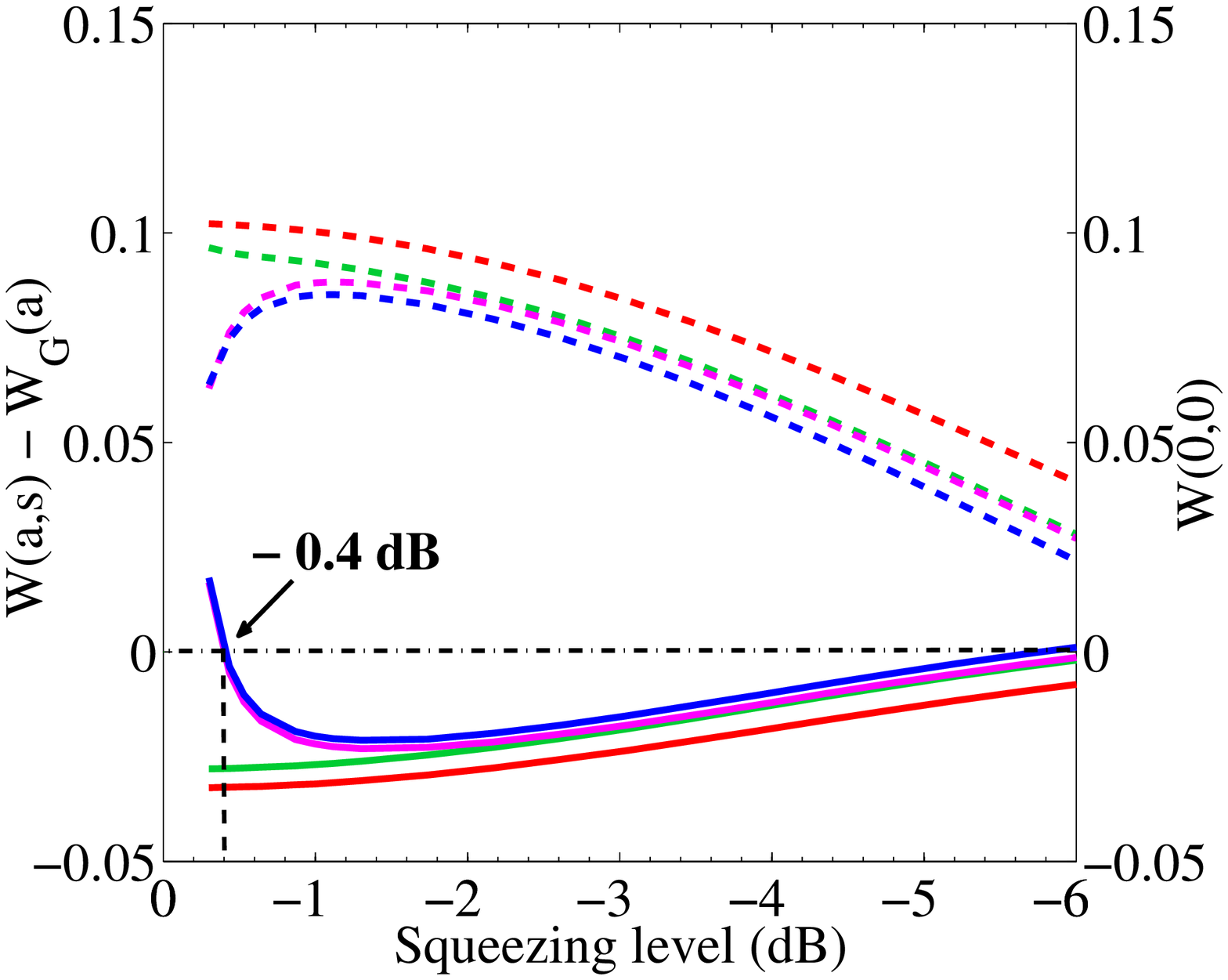}}
\caption{Quantum non-Gaussian character witness \& {\it W(0,0)} vs input state squeezing level, $V_{0}$, for (a) an InGaAs-APD and (b) a Si-APD. Dash lines and solid lines represent ${\it W(a,s) - W_G(a)}$ on left vertical axis and {\it W(0,0)} on right vertical axis, respectively. Red: PNRD, Green: NPNRD, Pink: IMPNRD, Blue: IMNPNRD. Pink and blue lines overlapped in (a). PNRD is a perfect photon-number-resolving detector and NPNRD is a perfect non-photon-number-resolving detector. IMPNRD is  an imperfect photon-number-resolving detector and IMNPNRD is an imperfect non-photon-number-resolving detector.}
\label{fig7squeezinglevel} 
\end{figure}

Figure \ref{fig7squeezinglevel} shows the variation of the quantum non-Gaussian character witness value and W(0,0) for a Schr\"{o}dinger kitten state with the squeezed vacuum state level, $V_{0}$ in dB, prepared with 1) perfect photon-number-resolving detector (PNRD), 2) a perfect non-photon-number-resolving detector (NPNRD), 3) an imperfect photon-number-resolving detector (IMPNRD), and 4) an imperfect non-photon-number-resolving detector (IMNPNRD). The other parameters used in these simulations to generate figure \ref{fig7squeezinglevel} are listed in table \ref{tableparamerters}. Figure \ref{fig7squeezinglevel}(a) shows that a negative Wigner function can not be observed for a Schr\"{o}dinger kitten state at telecommunication wavelengths prepared with an imperfect photon-number detector and the given experimental parameters. However, it is easy to obtain a quantum non-Gaussian state once the squeezing level (i.e. the minimum noise variance) exceeds $-$0.8 dB. On the contrary, when the squeezing level is as small as  $-0.4$ dB, it is possible to obtain a Schr\"{o}dinger kitten state with a negative Wigner function when prepared with a Si-APD , as shown in figure \ref{fig7squeezinglevel}(b).

\begin{figure} [htbp]
\centering
\subfigure[$P_{dc}=1\times10^{-4}$]{
\label{fig:subfig:a} 
\includegraphics[height=2.4in,width=2.95in]{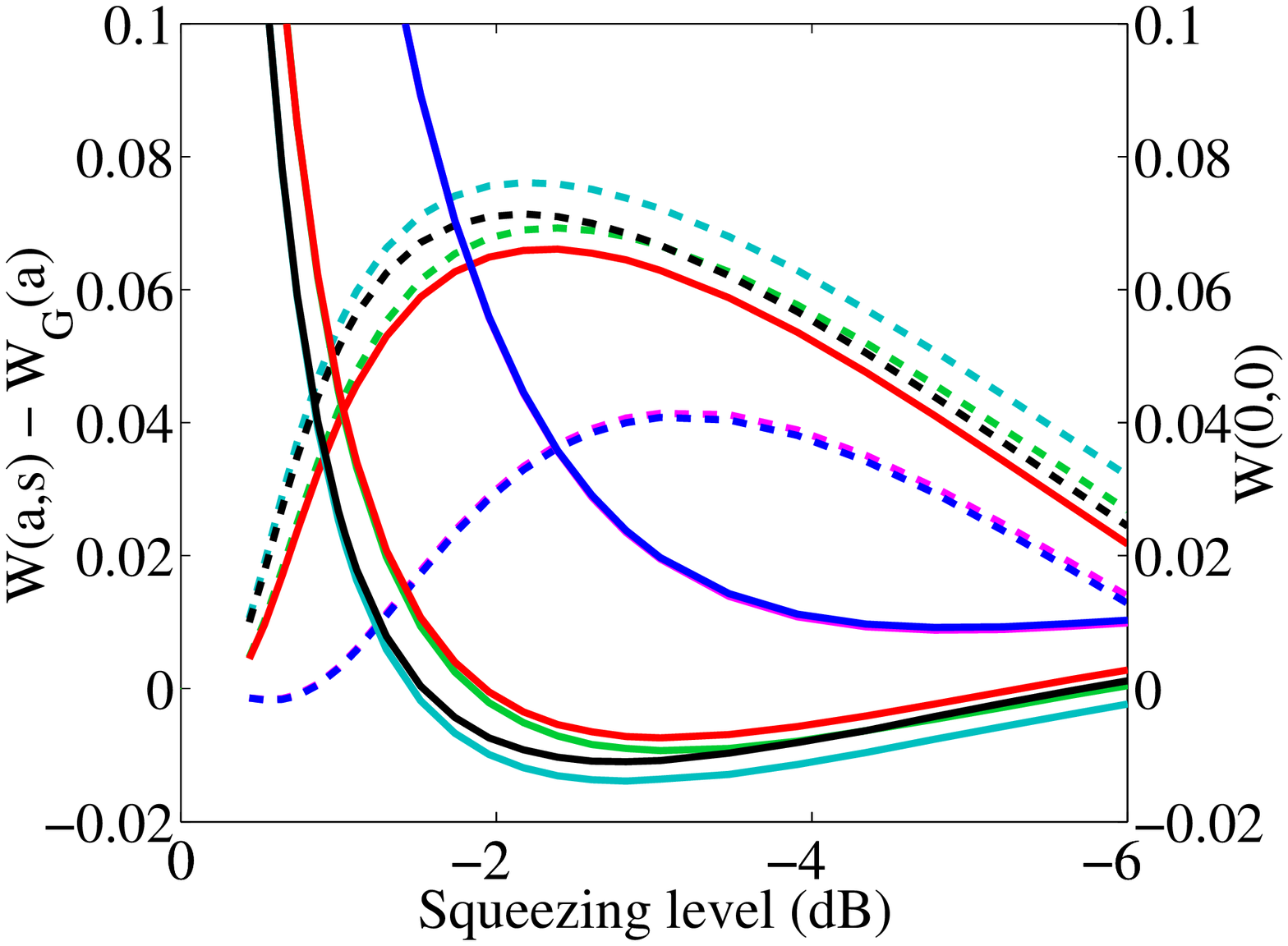}}
\hspace{0in}
\subfigure[$\eta_{APD}=10\%$]{ \label{fig:subfig:b} 
\includegraphics[height=2.4in,width=2.95in]{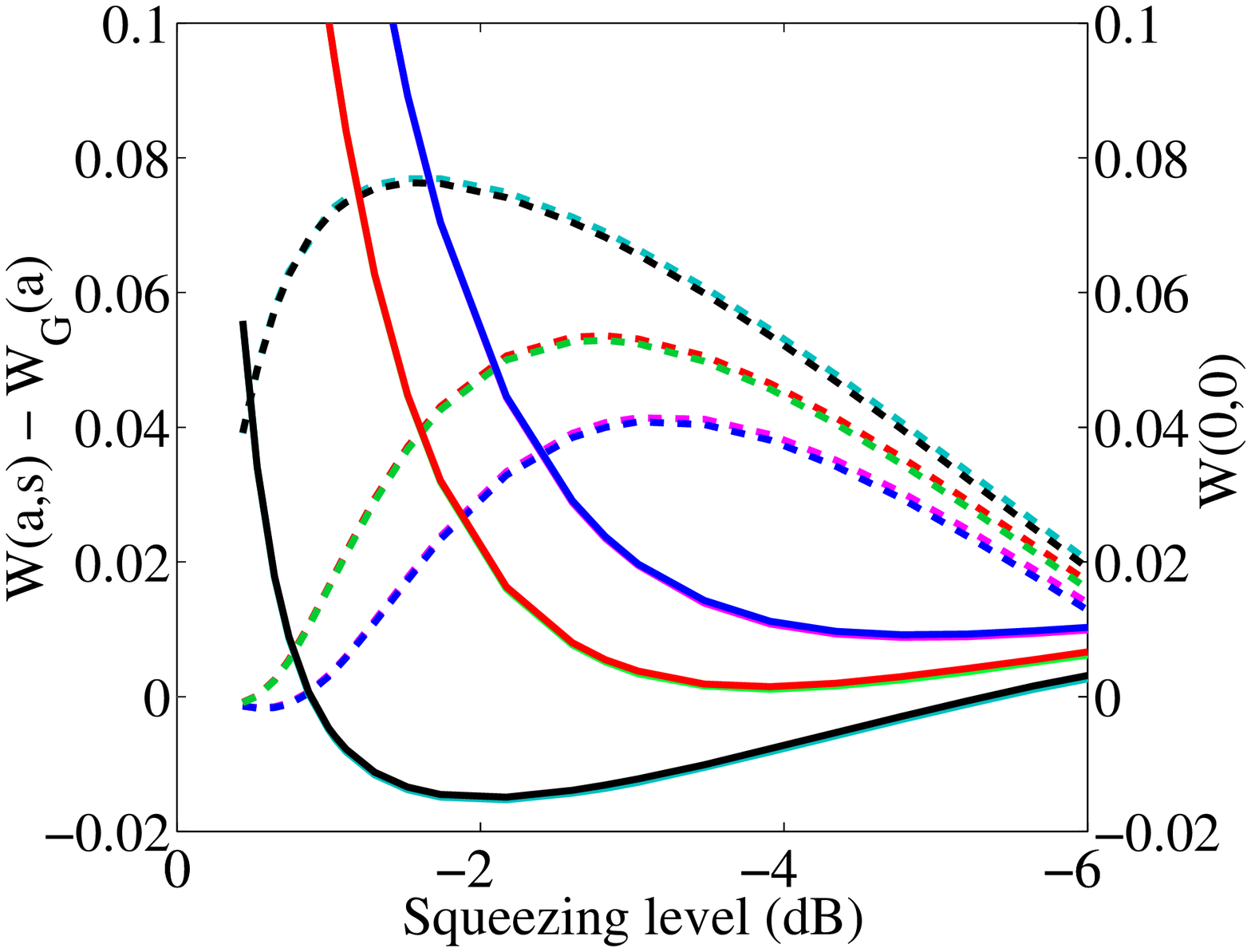}}
\caption{Squeezing level, $V_{0}$,impacts under (a) different detection efficiencies ($10\%$, $45\%$ and $65\%$ ) with $P_{dc}=1\times10^{-4}$ and (b) dark count probabilities ($1\times10^{-4}$, $5\times10^{-5}$, and $5\times10^{-6}$) with $\eta_{APD}=10\%$. Dash and solid lines represent ${\it W(a,s) - W_G(a)}$ on left vertical axis and {\it W(0,0)} on right vertical axis, respectively. Light blue: IMPNRD with (a) $\eta_{APD}$ = 65\% and (b) $P_{dc}$ = $5\times10^{-6}$, Black: IMNPNRD with (a)$\eta_{APD}$ = 65\% and (b) $P_{dc}$ = $5\times10^{-6}$, Green: IMPNRD with (a) $\eta_{APD}$ = 45\%  and (b) $P_{dc}$ = $5\times10^{-5}$, Red: IMNPNRD with (a) $\eta_{APD}$ = 45\% and (b) $P_{dc}$ = $5\times10^{-5}$, Pink: IMPNRD with (a) $\eta_{APD}$ = 10\% and (b) $P_{dc}$ = $1\times10^{-4}$, Blue: IMNPNRD with (a) $\eta_{APD}$ = 10\% and (b) $P_{dc}$ = $1\times10^{-4}$. }
\label{figsqueezinglevelwithpdcetaAPD} 
\end{figure}

In addition, it can be seen that there is an optimal value for the squeezing level of the input state to obtain maximal character witness value and minimal W(0,0). This is because when the squeezing level is lower than this optimal value, as shown in figure \ref{fig7squeezinglevel}(a), not enough photons are subtracted. Consequently, the dark counts will dominate the `real' click events. For large levels of squeezing (e.g. $-$6 dB), the probability to subtract more than one photon is dramatically increased. As a result, the one-photon- subtracted squeezed vacuum state is contaminated by two or three-photon-subtracted states. Therefore, the optimization of the input squeezing level is critical in the experiment design.

  Furthermore, a perfect PNRD demonstrates a significant advantage over a perfect NPNRD for both  the Si-APD and InGaAs-APD cases. However, such an advantage of the PNRD disappears in the case of the InGaAs-APD when imperfections, such as dark count and detection inefficiency, are taken into account.

   As both the dark count probability and detection efficiency of the InGaAs-APD are inferior to those of the Si-APD, we investigated the impacts of squeezing level for InGaAs-APDs with different detection efficiencies and dark count probabilities (see figure \ref{figsqueezinglevelwithpdcetaAPD}). Figure \ref{figsqueezinglevelwithpdcetaAPD}(a) implies that the advantage of an IMPNRD becomes distinguishable at higher squeezing levels when the APD detection efficiency is enhanced to 45\%. While very little difference is observed between an IMPNRD and an IMNPNRD in figure \ref{figsqueezinglevelwithpdcetaAPD}(b), despite the dark count probability being reduced to the same level as the Si-APD ($P_{dc}$ = 5$\times10^{-6}$). This reveals that the low detection efficiency of InGaAs-APDs substantially diminishes the advantage of an IMPNRD over an IMNPNRD.

  More importantly, reducing the dark count probability from $1\times10^{-4}$ to $5\times10^{-6}$ results in a significant decrease of W(0,0), which indicates that a lower dark count probability is {\it critical} to obtain negative Wigner function in Schr\"{o}dinger kitten state generation. Comparing figure \ref{figsqueezinglevelwithpdcetaAPD}(a) and (b), we can see that W(0,0) obtained from a TES with $P_{dc} = 1\times10^{-4}$ and ${\it \eta_{TES}} = 65 \% $ is similar to that from an InGaAs-APD with $P_{dc} = 5\times10^{-6}$ and ${\it \eta_{APD}}$ = 10 $\% $. This confirms the experimental result in reference \cite{NPHY2010}, and implies that a low dark count probability is more influential than a higher detection efficiency or photon-number-resolving ability in one-photon-subtracted squeezed state generation experiments at telecommunication wavelengths. Therefore, for an InGaAs-APD with adjustable detection efficiency and dark count probability , such as the detector id220, the setting with lowest dark count probability is preferable despite the smaller detection efficiency.

\begin{figure} [htbp]
\centering
\subfigure[InGaAs-APD]{
\label{fig:subfig:a} 
\includegraphics[height=2.4in,width=2.95in]{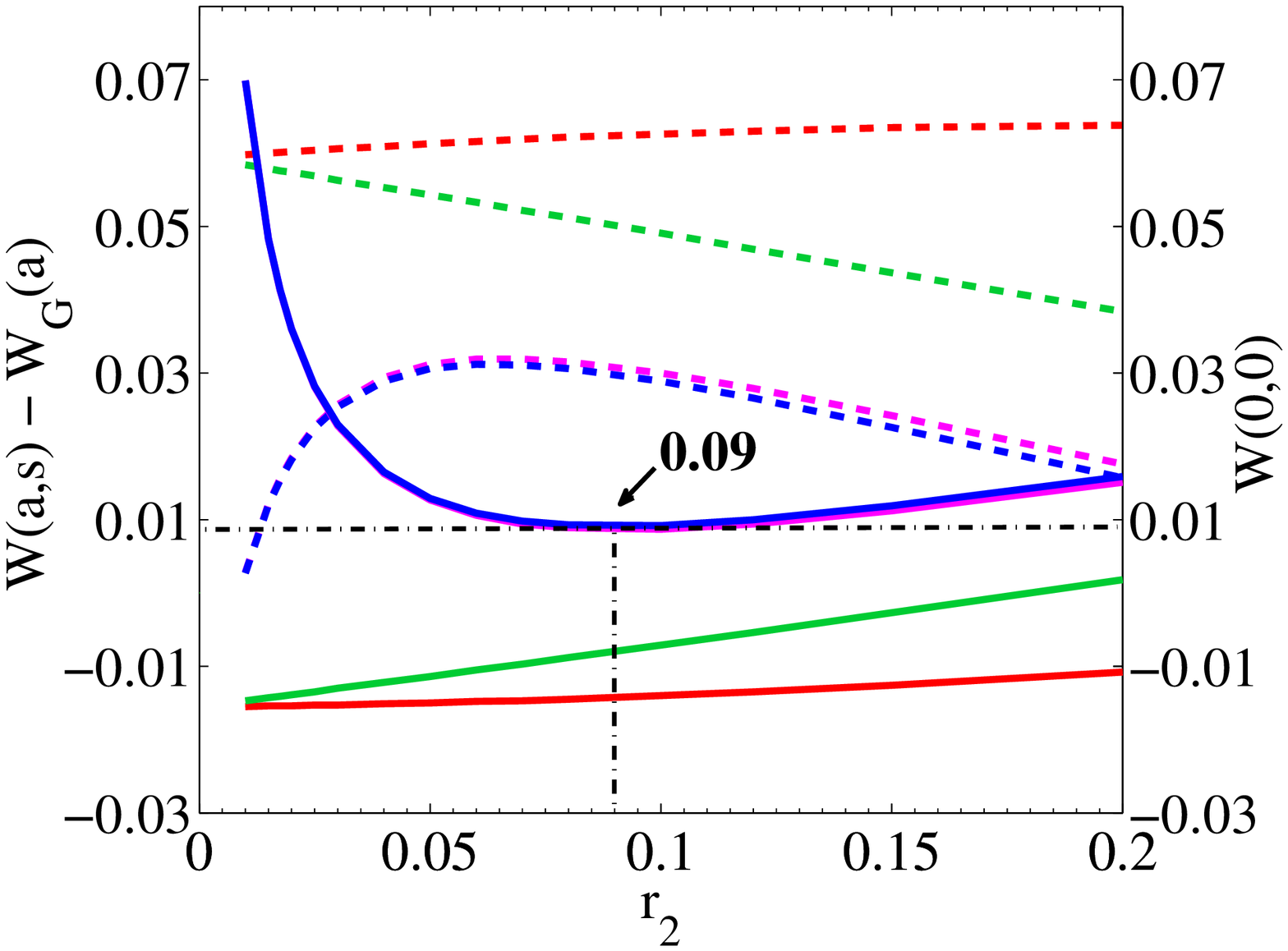}}
\hspace{0in}
\subfigure[Si-APD]{ \label{fig:subfig:b} 
\includegraphics[height=2.4in,width=2.95in]{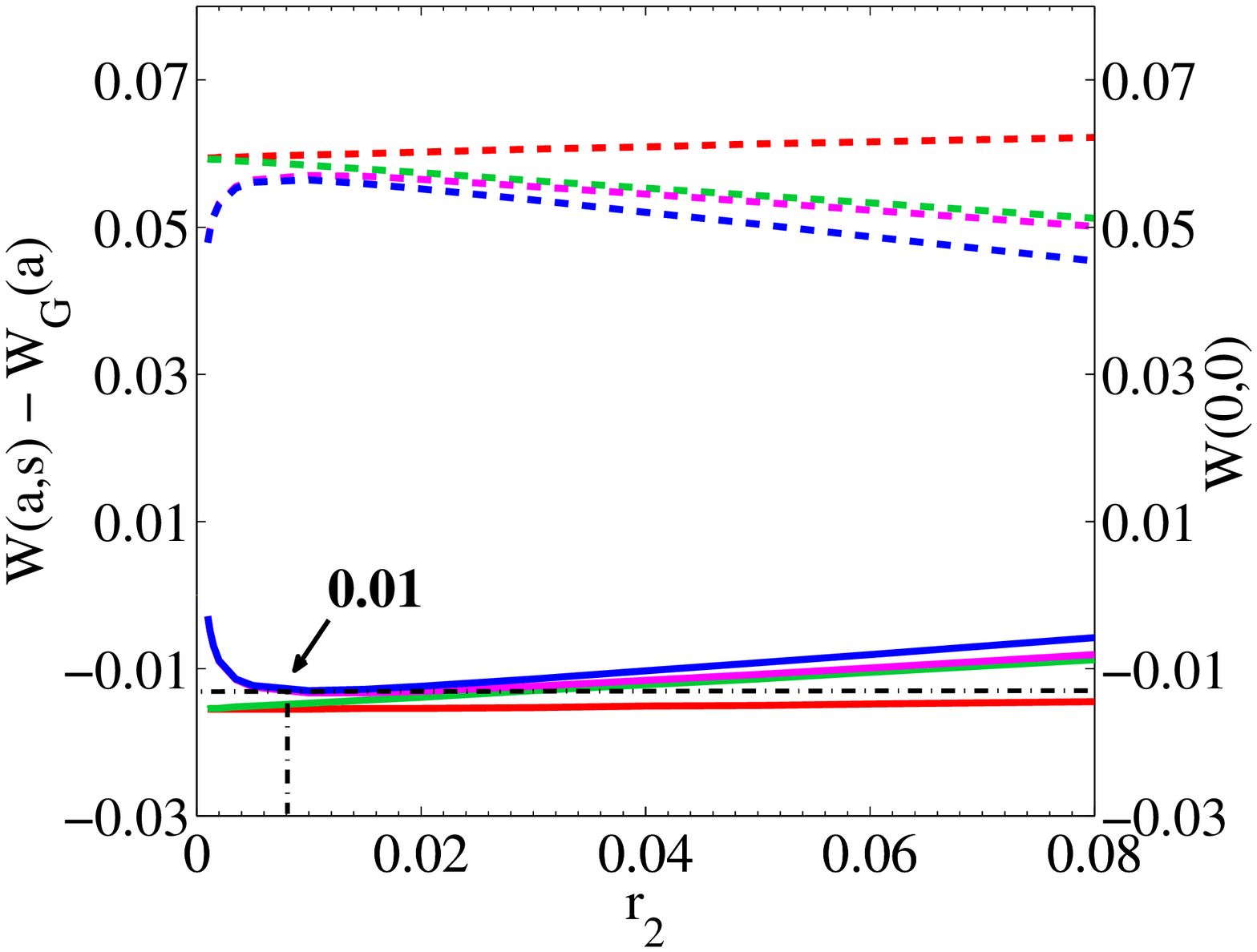}}
\caption{Quantum non-Gaussian character witness \& {\it W(0,0)} vs $r_{2}$ for (a) an InGaAs-APD and (b) a Si-APD. Dash lines and solid lines represent ${\it W(a,s) - W_G(a)}$ on left vertical axis and {\it W(0,0)} on right vertical axis, respectively. Red: PNRD, Green: NPNRD, Pink: IMPNRD, Blue: IMNPNRD. }
\label{figr2} 
\end{figure}


\begin{figure} [htpb]
\centering
\subfigure[InGaAs-APD]{
\label{fig:subfig:a} 
\includegraphics[height=2.4in,width=2.95in]{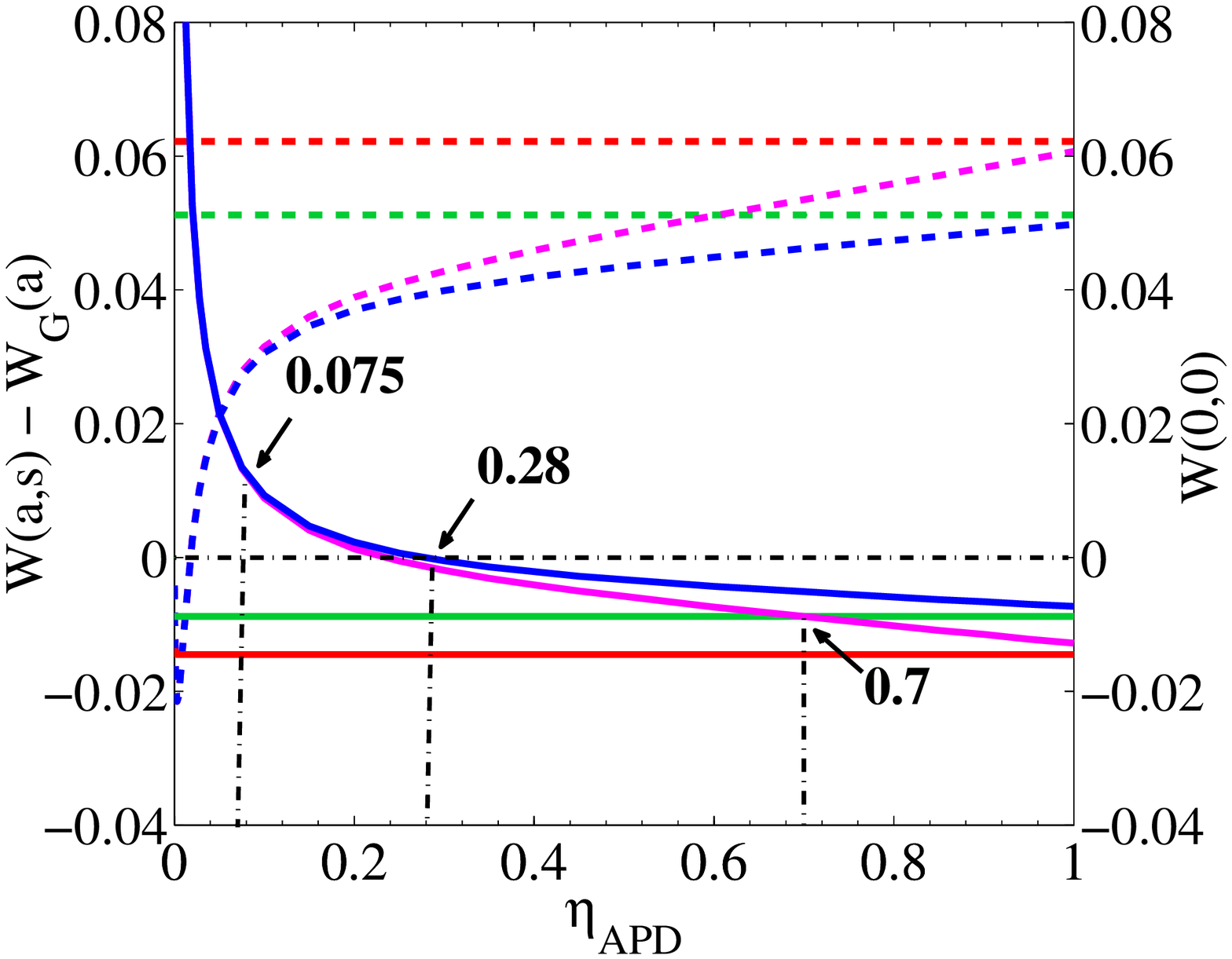}}
\hspace{0in}
\subfigure[Si-APD]{ \label{fig:subfig:b} 
\includegraphics[height=2.4in,width=2.95in]{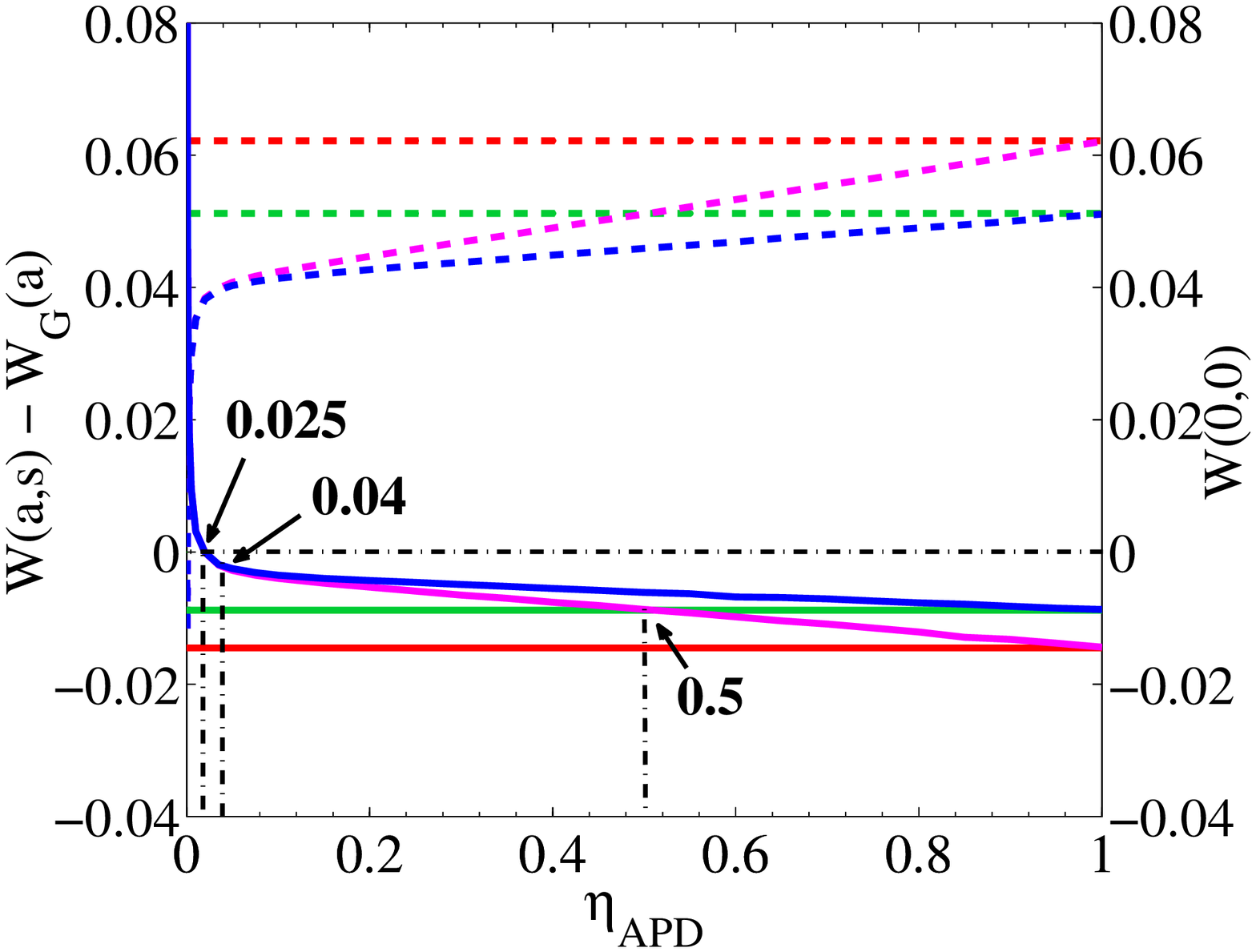}}
\caption{Quantum non-Gaussian character witness \& {\it W(0,0)} vs APD efficiency for (a) an InGaAs-APD and (b) a Si-APD. To obtain negative-valued Wigner function, teh requirements for the InGaAs-APD and Si-APD required to $\eta_{APD}$ $ >$ 28\%, and $\eta_{APD}$ $>$ 2.5\%, respectively. Dash lines and solid lines represent ${\it W(a,s) - W_G(a)}$ on left vertical axis and {\it W(0,0)} on right vertical axis, respectively.  Red: PNRD, Green: NPNRD, Pink: IMPNRD, Blue: IMNPNRD.  }
\label{figetaapd} 
\end{figure}

As discussed in table \ref{tablephysicalmechanism}, figure \ref{figr2} shows that the reflectivity, ${\it r_{2}}$, of the `magic' reflector has a similar impact to the input state squeezing level, $V_{0}$, due to the same physical mechanism. The reflectivity, $r_{2}$, must be optimized to obtain the maximum quantum non-Gaussian character witness value and minimum W(0,0). The optimal value of $r_{2}$ for a Schr\"{o}dinger kitten state prepared with a Si-APD ($r_{2opt}$=0.01) is notably smaller than that prepared with an InGaAs-APD ($r_{2opt}$=0.09). Under the circumstance of a Si-APD, the smaller $r_{2}$ results in a larger character witness value and a deeper W(0,0). However, if $r_{2}$ is too small, then it is easy to induce false clicks since the number of real APD counts, which are proportional to $r_{2}$, will be lower than the amount of dark counts. Therefore, it is necessary to compromise a small $r_{2}$ that is still higher enough to ensure the count rate is larger than the dark count rate. This has been validated by the results reported in most kitten state generation experiments using Si-APDs \cite {SCI2006, OE2007, PRA2006604, PRA2010}.

The effects of the APDs quantum efficiencies are shown in figure \ref{figetaapd}(a) and (b). As expected, the quantum efficiency has a similar impact on the quantum non-Gaussian character witness value and W(0,0) to squeezing level and $r_{2}$. It is noted that an IMPNRD does not demonstrate superiority to IMNPNRD until the detection efficiency increases to a specific value (for instance for the InGaAs-APD and Si-APD, $\eta_{APD}$ = 7.5\% and 4\%, respectively), which strengthens the argument that APD inefficiency and non-photon-number resolving ability give equivalent effects. Furthermore, when the detection efficiency is too low, the performance of an NPNRD is superior to an IMPNRD. This trend is due to the IMPNRD suffering from both inefficient detection and dark counts. However, when the detection efficiency is increased (for the InGaAs-APD and Si-APD, $\eta_{APD}$ = 70\% and 50\%, respectively), the impact from dark counts dominates in the IMPNRD. As a result, the performance of IMPNRD is gradually superior to that of the NPNRD, and approaches the performance of a perfect PNRD.

\subsubsection{Input state impurity, $r_{1}$, and homodyne detection efficiency, $\eta_{HD}$}

\begin{figure} [htbp]
\centering
\subfigure[InGaAs-APD]{
\label{fig:subfig:a} 
\includegraphics[height=2.4in,width=2.95in]{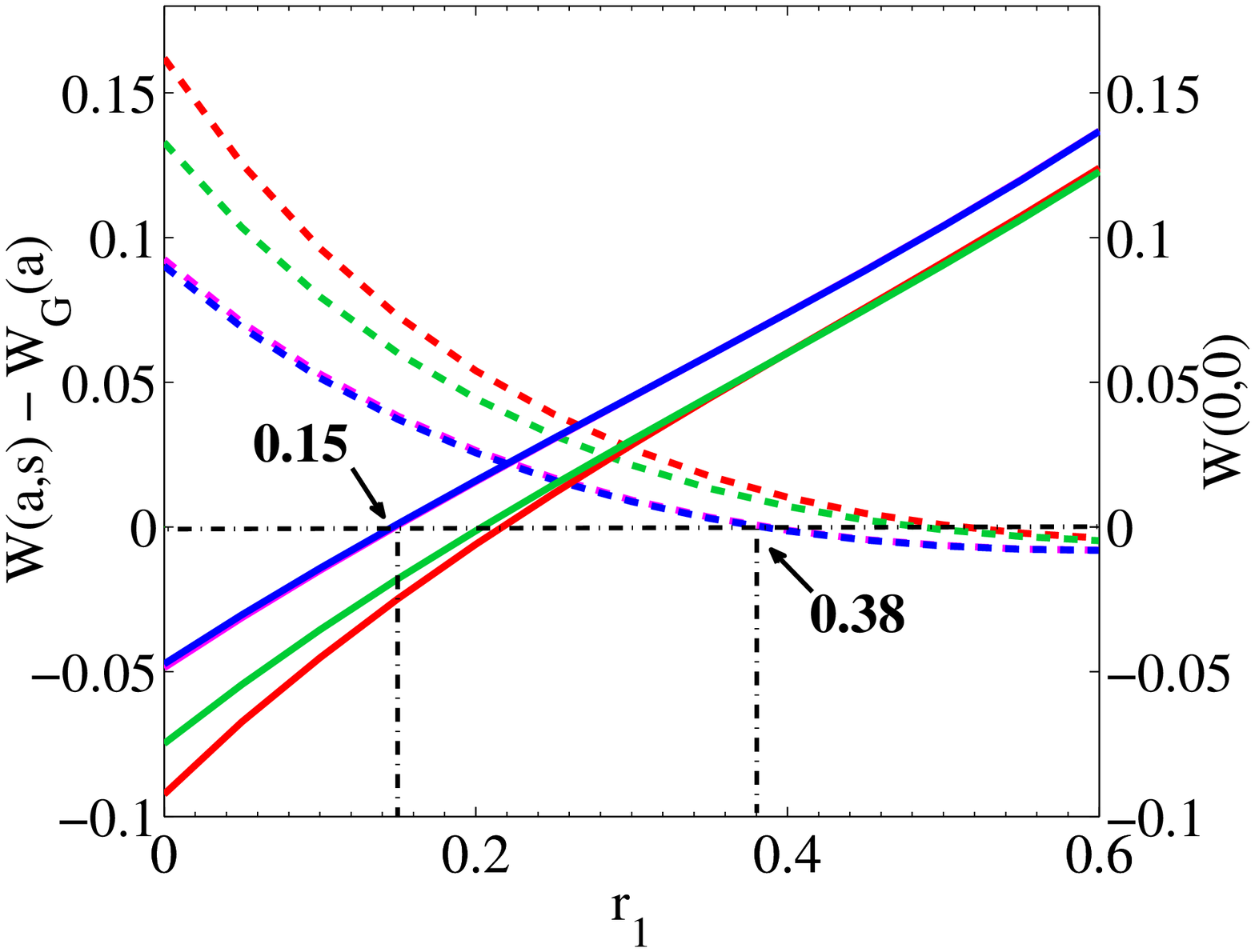}}
\hspace{0in}
\subfigure[Si-APD]{ \label{fig:subfig:b} 
\includegraphics[height=2.4in,width=2.95in]{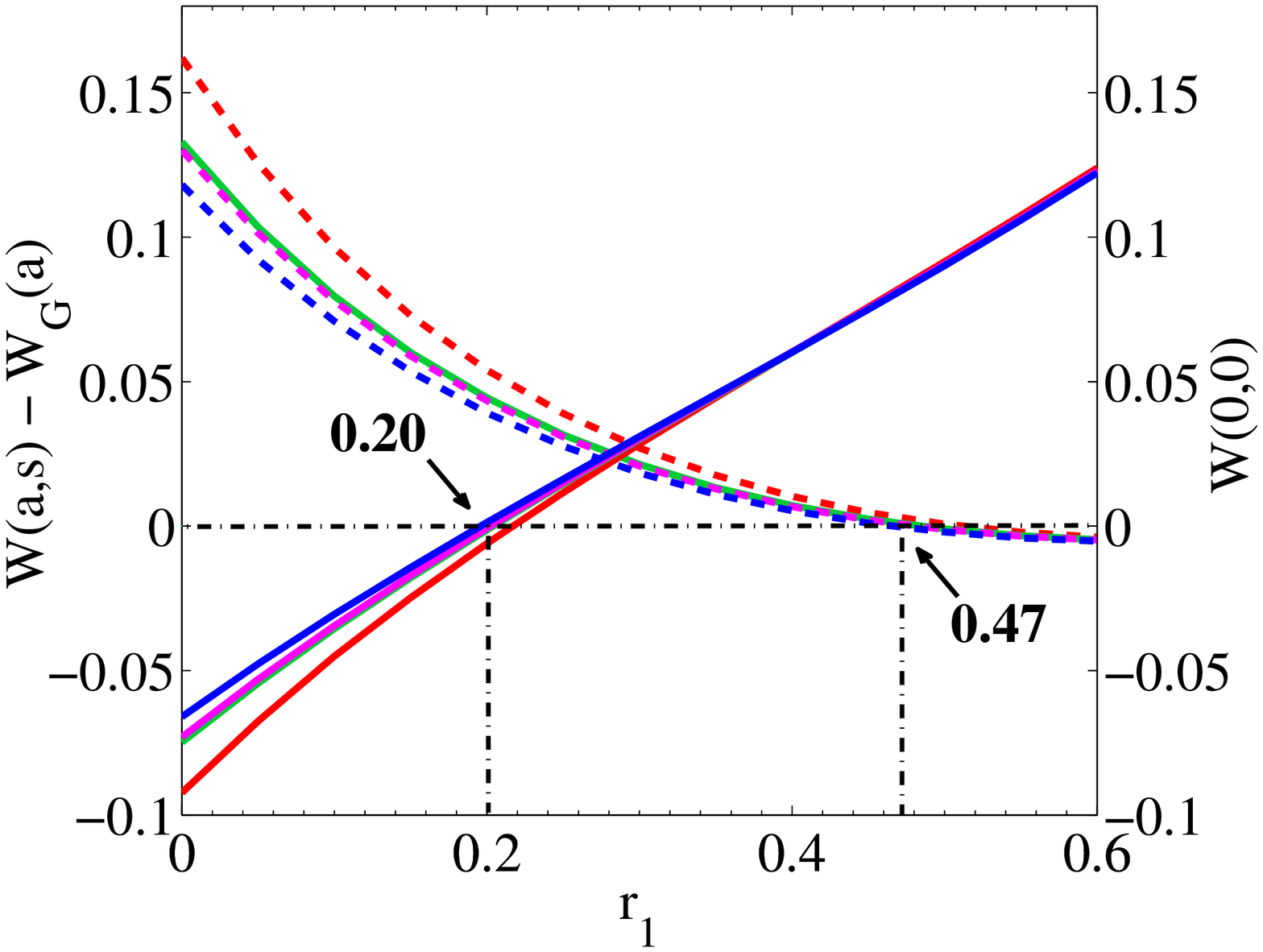}}
\caption{Quantum non-Gaussian character witness \& {\it W(0,0)} vs input state impurity $r_{1}$ for (a) an InGaAs-APD and (b) a Si-APD. To obtain negative-valued Wigner functions for the InGaAs-APD and the Si-APD, $r_{1}$ are required to be less than 0.15 and 0.20, respectively. The corresponding requirements to obtain quantum non-Gaussian states are $r_{1}$ $<$ 0.38 and $r_{1}$ $<$ 0.47 for the InGaAs-APD and the Si-APD, respectively. Dash lines and solid lines represent ${\it W(a,s) - W_G(a)}$ on left vertical axis and {\it W(0,0)} on right vertical axis, respectively.  Red: PNRD, Green: NPNRD, Pink: IMPNRD, Blue: IMNPNRD. Pinks and blue lines overlapped in (a). }
\label{figinputstateimpurity} 
\end{figure}

\begin{figure} [htbp]
\centering
\subfigure[InGaAs-APD]{
\label{fig:subfig:a} 
\includegraphics[height=2.4in,width=2.95in]{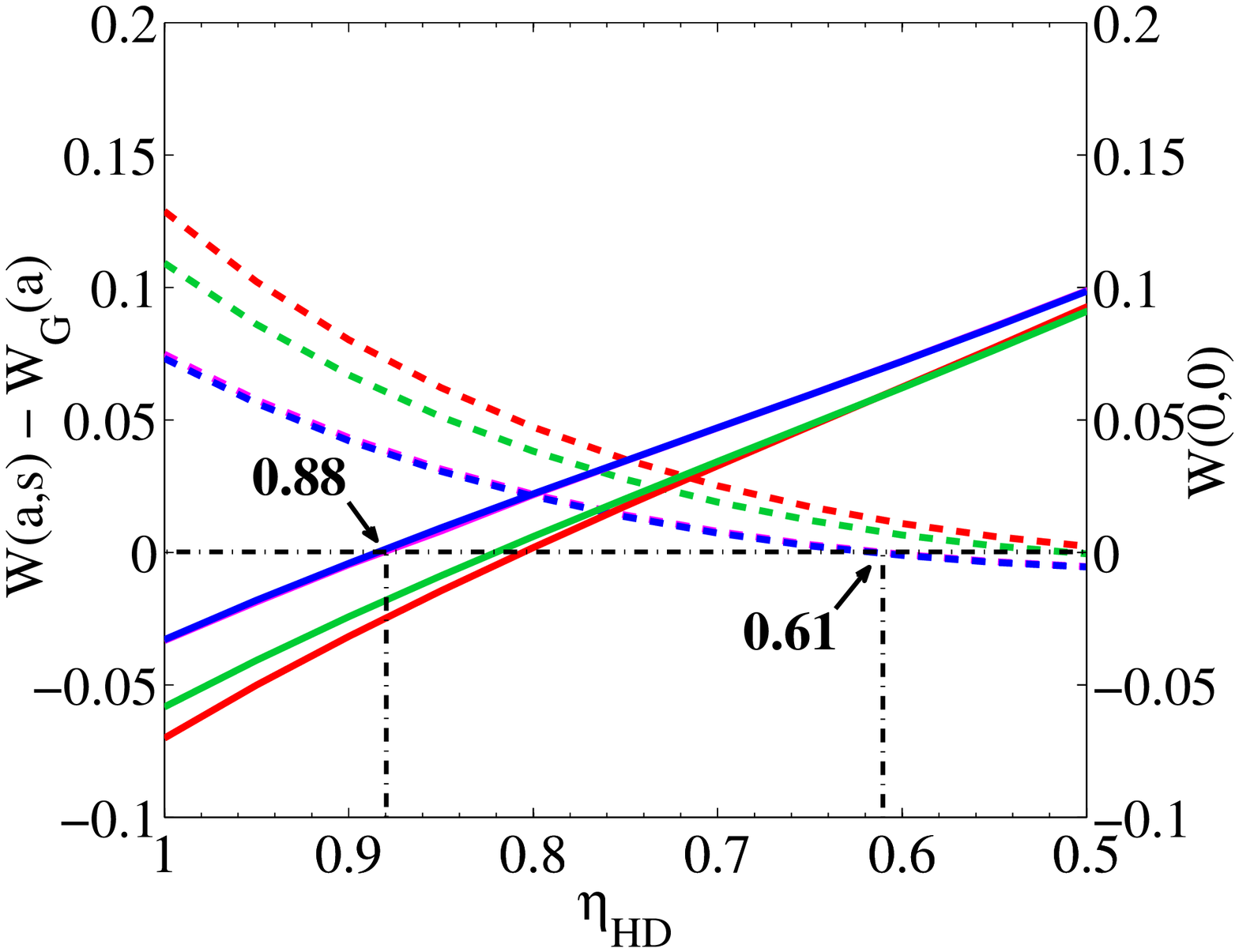}}
\hspace{0in}
\subfigure[Si-APD]{ \label{fig:subfig:b} 
\includegraphics[height=2.4in,width=2.95in]{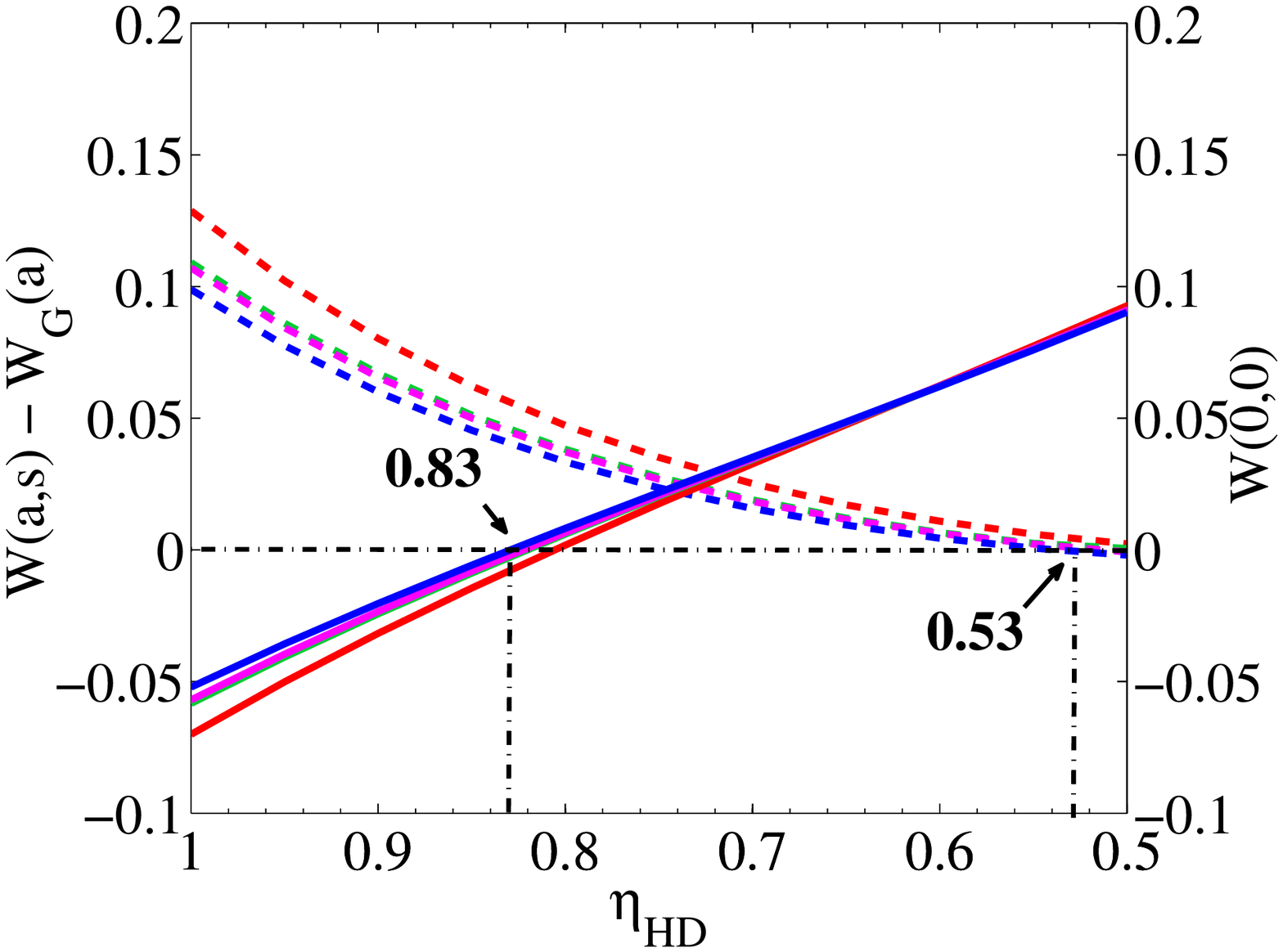}}
\caption{Quantum non-Gaussian character witness \& {\it W(0,0)} vs homodyne detection efficiency for (a) an InGaAs-APD and  (b) a Si-APD. To obtain negative-valued Wigner functions for the InGaAs-APD and the Si-APD, $\eta_{HD}$ are required to be higher than 0.88 and 0.83, respectively. The corresponding requirements to obtain quantum non-Gaussian states are $\eta_{HD}$ $>$ 0.61 and $\eta_{HD}$ $>$ 0.53 for the InGaAs-APD and the Si-APD, respectively. Dash lines and solid lines represent ${\it W(a,s) - W_G(a)}$ on left vertical axis and {\it W(0,0)} on right vertical axis, respectively.  Red: PNRD, Green: NPNRD, Pink: IMPNRD, Blue: IMNPNRD. Pink and blue lines overlapped in (a).}
\label{figetahd} 
\end{figure}

As shown in figures \ref{figinputstateimpurity} and \ref{figetahd}, the same physical mechanism underlying the input state impurity and homodyne detection inefficiency in Schr\"{o}dinger kitten state generation results in similar quantitative impacts on the quantum non-Gaussian character witness and W(0,0). The superior performance of the Si-APD to the InGaAs-APD culminates in lower requirements on the purity of the input state and homodyne detection efficiency at $\sim$ 860 nm kitten generation experiments compared to such states generation at $\sim$ 1550 nm.

\subsubsection{Dark count probability of a photon-number detector, $P_{dc}$, and mode impurity}
\begin{figure} [htpb]
\centering
\subfigure[InGaAs-APD]{
\label{fig:subfig:a} 
\includegraphics[height=2.4in,width=2.95in]{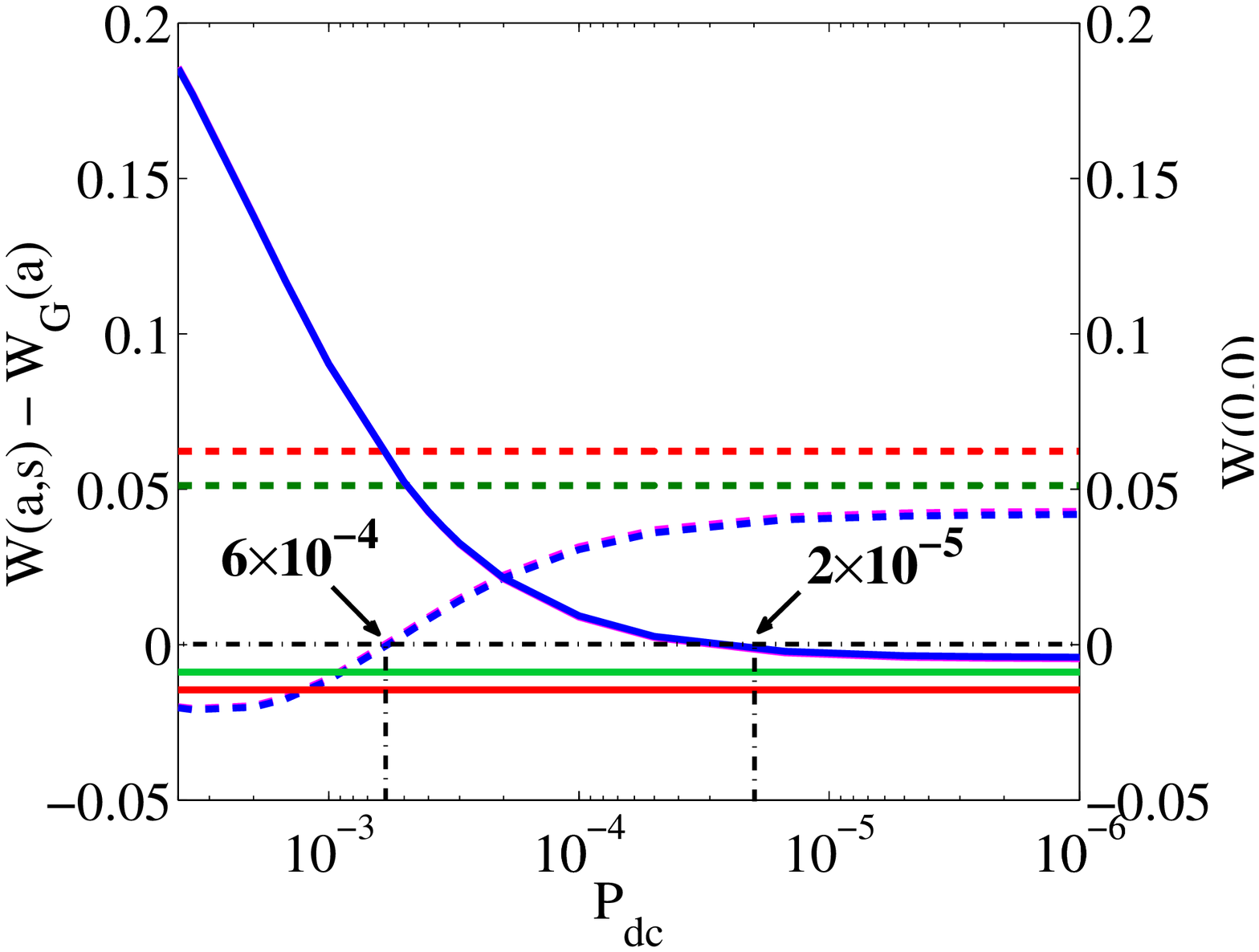}}
\hspace{0in}
\subfigure[Si-APD]{ \label{fig:subfig:b} 
\includegraphics[height=2.4in,width=2.95in]{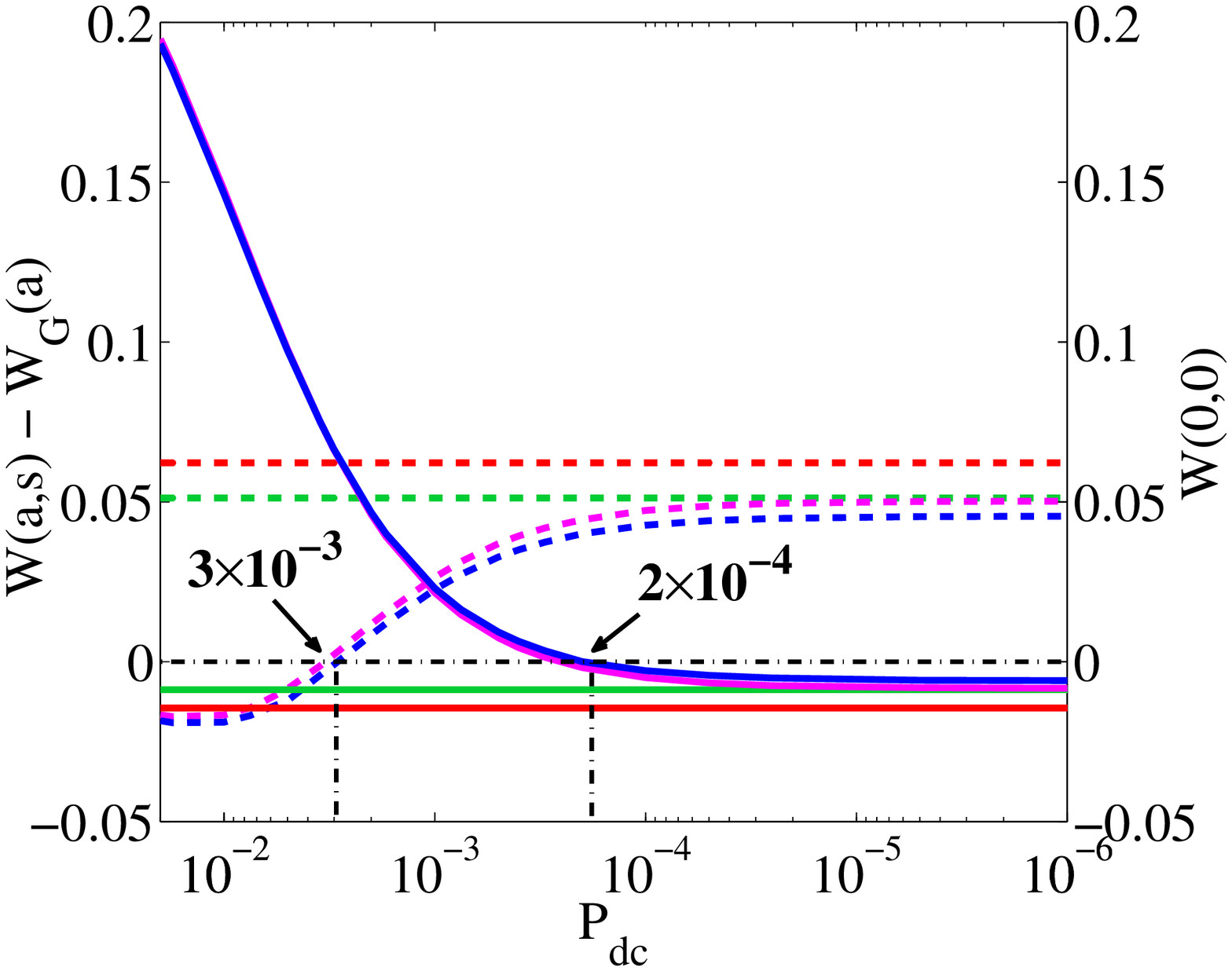}}
\caption{Quantum non-Gaussian character witness \& {\it W(0,0)} vs dark count for (a) an InGaAs-APD and (b) a Si-APD. To obtain quantum non-Gaussian states, $P_{dc}$ are required to be less than $6\times10^{-4}$ and $3\times10^{-3}$, for the InGaAs-APD and the Si-APD, respectively. Dash lines and solid lines represent ${\it W(a,s) - W_G(a)}$ on left vertical axis and {\it W(0,0)} on right vertical axis, respectively.  Red: PNRD, Green: NPNRD, Pink: IMPNRD, Blue: IMNPNRD. Pink and blue lines overlapped in (a).}
\label{figPdc} 
\end{figure}

Both dark count probability and mode impurity cause `false' clicks (i.e. click event is recorded even if no photon is actually subtracted).
As shown in figure \ref{figPdc}, to obtain W(0,0) $<$ 0, the dark count probability of an InGaAs-APD is required to be less than $2\times10^{-5}$, which is one order of magnitude {\it lower} than that of the Si-APD ($P_{dc}$ $<$ $2\times10^{-4}$) due to the lower detection efficiency of the InGaAs-APD. However, the dark count probabilities of most commercially available photon-number detectors for 1550 nm are far larger than that for 860 nm, as shown in table \ref{tableparamerters}. This gives a sound reason as to why it is difficult to obtain negativity in the Wigner function for Schr\"{o}dinger kitten states at telecommunication wavelengths.

\begin{figure} [htpb]
\centering
\subfigure[InGaAs-APD]{
\label{fig:subfig:a} 
\includegraphics[height=2.4in,width=2.95in]{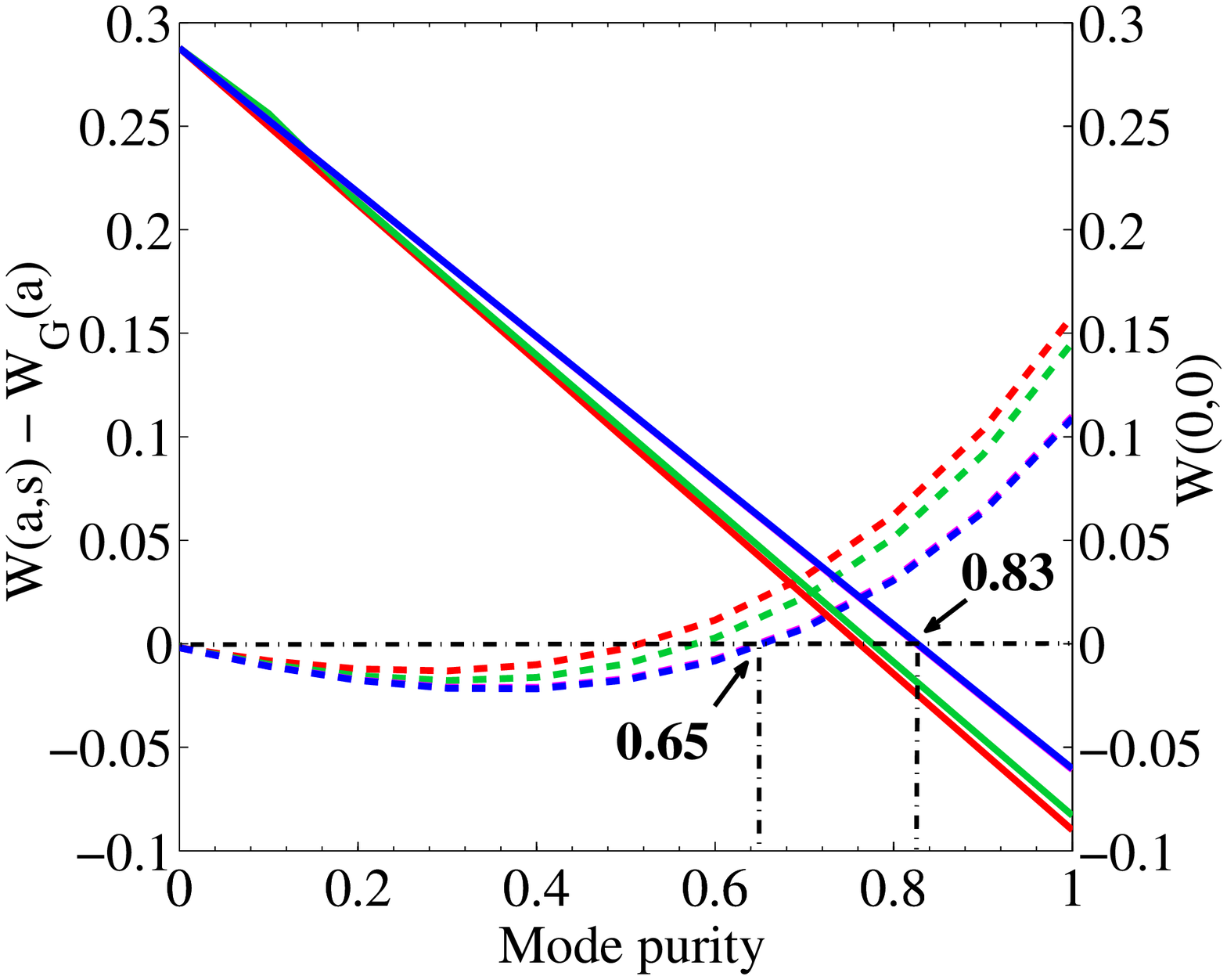}}
\hspace{0in}
\subfigure[Si-APD]{ \label{fig:subfig:b} 
\includegraphics[height=2.4in,width=2.95in]{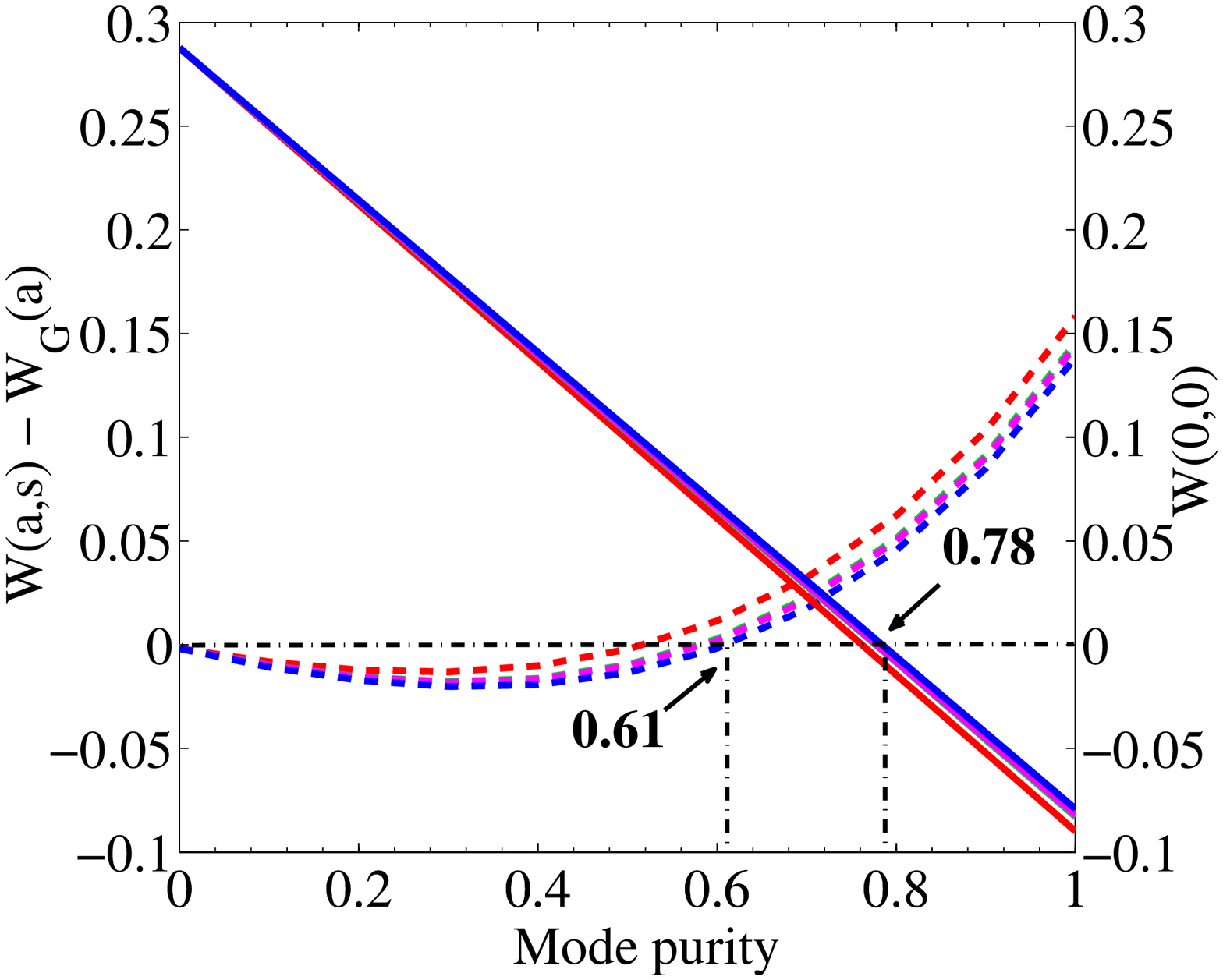}}
\caption{Quantum non-Gaussian character witness \& {\it W(0,0)} vs modal purity for (a) an InGaAs-APD and (b) a Si-APD.  To obtain negative-valued Wigner functions for the InGaAs-APD and the Si-APD, mode purity are required to be higher than 0.83 and 0.78, respectively. The corresponding requirements for mode purity to obtain quantum non-Gaussian states are higher than 0.65 and 0.61, for the InGaAs-APD and the Si-APD, respectively. Dash lines and solid lines represent ${\it W(a,s) - W_G(a)}$ on left vertical axis and {\it W(0,0)} on right vertical axis, respectively.  Red: PNRD, Green: NPNRD, Pink: IMPNRD, Blue: IMNPNRD. Pink and blue lines overlapped in (a). Pink, blue and green lines are quite close in (b).}
\label{figmodepurity} 
\end{figure}

The change seen in the character witness and W(0,0) for varying mode purity is shown in figure \ref{figmodepurity}.  The PNRD is superior to the NPNRD when the mode purity is high, but the advantage of PNRD gradually declines when the mode purity is too low to successfully projected the Schr\"{o}dinger kitten state.

\section{Conclusions}

We quantitatively analysed the impacts of a full set of experimental imperfections on Schr\"{o}dinger kitten state generation in terms of the quantum non-Gaussian character witness and Wigner function. According to the comparison between Schr\"{o}dinger kitten states prepared with an InGaAs-APD and a Si-APD, the inferiority of telecommunication-wavelength photon-number detectors justifies the higher requirements on the optical experimental parameters to obtain negativity in the Wigner function. Furthermore, the lower detection efficiency of commercially available photon-number detectors dramatically degrades the superiority of the photon-number-resolving detector for one-photon projected Schr\"{o}dinger kitten state generation at telecommunication wavelengths. The dark count probability of InGaAs-APDs is required to be on the order of $10^{-5}$ to obtain negative values at W(0,0). This discussion on the effects of various experimental parameters guides the analysis of kitten state generation experiments for particular wavelengths. It is clear that Schr\"{o}dinger kitten state generation at telecommunication wavelengths presents numerous challenges but we can overcome these obstacles with thoughtful planning and careful experimental design.

\section{Acknowledgements}

Hongbin Song would like to thank Dr. Guofeng Zhang for helpful discussions. This work was supported financially by the Australian Research Council projects CE110001029 and DP1094650.

\section{Referencing}

\end{document}